\documentclass[iop]{emulateapj}

\usepackage {pstricks}

\shorttitle{Beyond MLT}
\shortauthors{Arnett  et al.}

\newcommand{\etal} { et~al.\ }  

\def \nuc#1#2{\relax\ifmmode{}^{#1}{\protect\text{#2}}\else${}^{#1}$#2\fi}

\begin{document}

\title{Beyond Mixing-length Theory: a step toward 321D}

\author{W. David Arnett\altaffilmark{1,2}
%\email{ wdarnett@gmail.com}
}
\author{Casey Meakin\altaffilmark{1,5}
%\email{casey.meakin@gmail.com }
}
\author{Maxime Viallet\altaffilmark{3}
%\email{mviallet@mpa-garching.mpg.de}
}
\author{Simon W. Campbell\altaffilmark{4,3}
%\email{simon.campbell@monash.edu}
}
\author{John C. Lattanzio\altaffilmark{4}
%\email{john.lattanzio@monash.edu}
}
\author{Miroslav Mo\'cak\altaffilmark{5}
%\email{john.lattanzio@monash.edu}
}

\altaffiltext{1}{Steward Observatory, University of Arizona, 
933 N. Cherry Avenue, Tucson AZ 85721}

%\altaffiltext{2}{ICRAnet, Rome, Pescara, Nice}
\altaffiltext{2}{Aspen Center for Physics, Aspen CO 81611}

%\altaffiltext{3}{Kavli Institute of Theoretical Physics, University of
%California, Santa Barbara, CA}

\altaffiltext{3}{Max-Planck Institut f\"ur Astrophysik, Garching, Deutschland}

\altaffiltext{4}{Monash University, Clayton, Victoria, Australia}

\altaffiltext{5}{Theoretical Division, LANL,
Los Alamos NM 87545}

\begin{abstract}

We examine the physical basis for algorithms to replace mixing-length theory (MLT) in stellar evolutionary computations. Our 321D procedure is based on numerical solutions of the Navier-Stokes equations. These  implicit large eddy simulations (ILES) are three-dimensional (3D), time-dependent, and turbulent,  including the Kolmogorov cascade.
We use the Reynolds-averaged Navier-Stokes (RANS)  formulation to make concise the 3D simulation data, and use the 3D simulations to give closure for the RANS equations.
We further analyze this data set with a simple analytical model, which is non-local and time-dependent, and which contains both MLT and the Lorenz convective roll as particular subsets of  solutions.
A characteristic length (the damping length) again emerges in the simulations; it is determined by an observed balance between (1) the large-scale driving, and (2) small-scale damping. 

The nature of mixing and convective boundaries is analyzed, including dynamic, thermal and compositional effects, and compared to a simple model.
 We find that
 (1) braking regions  (boundary layers in which mixing occurs) automatically appear {\it beyond} the edges of convection as defined by the Schwarzschild criterion, 
 (2) dynamic (non-local) terms imply a non-zero turbulent kinetic energy flux (unlike MLT),
 (3) the effects of composition gradients on flow can be comparable to thermal effects,
and (4) convective boundaries in neutrino-cooled stages differ in nature from those in photon-cooled stages (different P\'eclet numbers).
The algorithms are based upon ILES solutions to the Navier-Stokes equations, so that, unlike MLT, they do not require any calibration to astronomical systems in order to predict stellar properties. Implications for solar abundances, helioseismology, asteroseismology, nucleosynthesis yields, supernova progenitors and core collapse are indicated.

%{\bf DRAFT FROM \today}
\end{abstract}


\section{Introduction}

{\it Make everything as simple as possible, but no simpler.}  -Albert Einstein.\footnote{ 
This phrasing has often been attributed to Einstein, but might have originated as a verbal quip rather than in written text. For a discussion see \url{http://quoteinvestigator.com/2011/05/13/einstein-simple}.
} 

\

Stars contain three dimensional (3D), turbulent plasma. They are much more complex than the simplified one dimensional (1D) models we use for stellar evolution. Computer power is not adequate\footnote{See \cite{herwig14} as an example of the state of the art.} at present for well-resolved (i.e., turbulent) 3D simulations of {\it whole} stars for {\it evolutionary} timescales.

We attempt to tame this complexity by (1) use of 3D simulations as a foundation, 
(2) application of the Reynolds-Averaged Navier-Stokes (RANS) procedure \citep{ma07b,vmam13} to these simulations to discover dominant terms (closing the RANS system), and (3) construction of simple physical models, consistent with the 3D simulations, for use in stellar evolution codes. We call 
this approach ``321D'' because a central feature is the projection of 3D simulations down to 1D for 
use as a replacement for mixing-length theory (MLT; \citealt{bv58}). The process is designed to allow testing, extension, and systematic improvement.

Formally, the RANS equations are incomplete unless taken to infinite order\footnote{This occurs because the momentum equation is nonlinear, 
so  that each level of correlation requires the next higher level for its solution \citep{tritton,vallis}, giving an infinite regression.  See also \cite{cubarsi}.
}; they must be {\em closed} by truncation at low order to be useful. This need for truncation is due to the nature of the Reynolds averaging, which allows {\em all} fluctuations rather than only {\em dynamically consistent} ones.  Closure requires additional information to remove these extraneous solutions. Using 3D simulations avoids this problem by providing only dynamically consistent fluctuations.

As a complement to the full RANS approach, we consider approximations which focus on dynamics; these  provide a connection to historical work on convection in astrophysics and meteorology. 
Such a minimalist step may be easier to implement in stellar evolutionary codes, and still provide physical insight.
In the turbulent cascade, kinetic energy and momentum are concentrated in the largest eddies. Our approximate model contains both the largest eddies and the Kolmogorov cascade.

\subsection{Historical Background}

Erika B\"ohm-Vitense developed the version of mixing-length theory used in stellar evolution in the 1950s \citep{vitense53,bv58}, prior to the publication in the west of Andrey Kolmogorov's theory of the turbulent cascade \citep{kolmg}. MLT might have been different had she been aware of the original work \citep{kolmg41}. Edward Lorenz  showed that a simple convective roll had chaotic behavior (a strange attractor, \citealt{lorenz}). Ludwig Prandtl developed the theory of boundary layers \citep{prandtl}, as well as the original version of MLT \citep{prandtl25}. All these ideas will be relevant to our discussion, which is based, as far as possible, upon experimentally verified turbulence theory and 3D simulations, and free of astronomical calibration.

\begin{figure}[h]
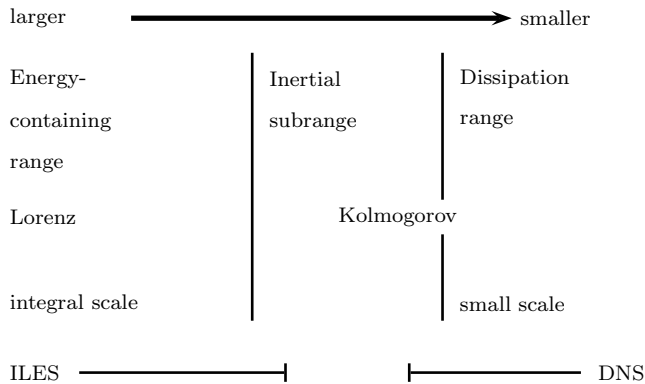

\figurenum{1}
\psset{unit=2.3cm}
\pspicture*[](-1.5,-1.7)(2.5,0.7)
\rput*[l]{0}(-1.4,0.1){Energy-}
\rput*[l]{0}(-1.4,-0.15){containing}
\rput*[l]{0}(-1.4,-0.4){range}
\rput*[l]{0}(-1.4,-0.7){Lorenz}

\psline[linewidth=1pt]{-}(-0.,0.25)(-0.,-1.3)

\rput*[l]{0}(0.1,0.1){Inertial}
\rput*[l]{0}(0.1,-0.15){subrange}

\psline[linewidth=1pt]{-}(1.1,0.25)(1.1,-1.3)

\rput*[l]{0}(1.2,0.1){Dissipation}
\rput*[l]{0}(1.2,-0.15){range}
\rput*[l]{0}(0.5,-0.7){Kolmogorov}

\psline[linewidth=2pt]{->}(-0.7,0.45)(1.5,0.45)
\rput*[l]{0}(-1.4,0.45){larger}
\rput*[l]{0}(1.55,0.45){smaller}

\rput*[l]{0}(-1.4,-1.2){integral scale}
\rput*[l]{0}(1.2,-1.2){small scale}

\psline[linewidth=1pt]{-|}(-1.0,-1.6)(0.2,-1.6)
\rput*[l]{0}(-1.4,-1.6){ILES}
\psline[linewidth=1pt]{|-}(0.9,-1.6)(1.9,-1.6)
\rput*[l]{0}(2.0,-1.6){DNS}

\endpspicture

\caption{The 3D turbulent energy cascade on a logarithmic scale of sizes; see \cite{davidson,pope}. The arrow indicates the direction of net energy flow. The range of applicability of Implicit Large Eddy Simulations (ILES) and Direct Numerical Simulation (DNS) are shown. 
See text for discussion.}
\label{cascade}
\end{figure}
\placefigure{1}

The 3D turbulent energy cascade is illustrated in Figure~\ref{cascade}. The turbulent motion is driven at the largest scale (the ``integral'' scale), which contains most of the kinetic energy. These motions are unstable and break up into smaller-scale flow patterns dominated by inertial forces (the ``inertial subrange"). This continues to scales small enough for microscopic effects (viscosity) to finally provide damping of the flow at the Kolmogorov scale. Both the inertial subrange and the dissipation range are insensitive to the details of the boundary conditions at the integral scale, and are ``universal'' in this sense.  We use the term ``universality" to mean the property of insensitivity to boundary conditions at the integral scale.
\cite{kolmg41}  found the striking result that the rate of dissipation is insensitive to the value of the viscosity, but is determined by the rate that the largest-scale flows feed the cascade. {\em This behavior of the non-linear flow ``hides'' the microscopic value of the viscosity.} We use Kolmogorov theory to describe the flow in the range where universality holds.

Direct Numerical Simulations (DNS) resolve the small  scales at which dissipation happens, and can extend up to the inertial range, but not to stellar scales. Implicit Large Eddy Simulations (ILES) can extend from stellar (integral) scales down to the inertial range, but not to the dissipation range. Fig.~\ref{cascade} illustrates both.
%
%\begin{deluxetable*}{lllll}
%\tablewidth{440pt}
%\tablecaption{Turbulence approximations (increasing simplicity)\label{table1}}
%\tabletypesize{\small}
%
%\tablehead{ \colhead{level} & \colhead{reference} &\colhead{complexity} & \colhead{code} 
%& \colhead{time dependence}   
%}
%\startdata
%Real stars & --- & $(\sim 10^{48})^{a}$  &--- &--- \\
%\\
%3D ILES & \cite{herwig14} &$(\sim 10^{10})^{b}$ & PPB &  chaotic\\
%3D ILES & \cite{ma07b} &$(\sim 10^{9})^{b}$ & PROMPI &  chaotic\\
% & \cite{maxime} & & MUSIC & chaotic \\
%\\
%RANS & \cite{ma07b} & $\sim 20$, $\geq 3$ ode$^c$ & --- & turnover and \\
% & \cite{vmam13} & &  & angular averages  \\
%\\
%roll & \cite{lorenz} & 3 ode & --- &  chaotic\\
% & \cite{am11b} & single mode  & & chaotic \\
%\\
%321D$^{d}$ & this paper & +1 ode & TYCHO$^{e}$ & cell average\\
%\\
%MLT & \cite{bv58} & algebraic & (MESA,  & steady-state \\
% & \cite{kippen} & (cubic equation) & MONSTAR, &  \\
%  & \cite{sa14} & &GENEC, etc.) &  \\
%
%\enddata
%\tablenotetext{a}{Turbulent degrees of freedom in the Sun; see \cite{amv14} for details of estimate.}
%\tablenotetext{b}{Turbulent degrees of freedom available on grid (ILES).}
%\tablenotetext{c}{ode: ordinary differential equation; dependent on the nuclear reaction network used, in general.}
%\tablenotetext{d}{Here we take a first step. See \S\ref{dynamics} for references to some other attempts to improve MLT.}
%\tablenotetext{e}{321D implementations are planned in MESA, MONSTAR and GENEC; 321D will be generally available.}
%
%\end{deluxetable*}
%
%\placetable{1}
%

Landau objected to the notion of complete universality on the grounds that the largest scales were subject to boundary conditions which would be specific to the case in question \citep{ll59,frisch}. We will  incorporate this idea by splitting the turbulent flow into two parts: the integral-scale motion and the turbulent cascade.
As an aid to understanding the integrated properties of the integral-scale motion, we are guided by the simplest model of a convective roll, due to \cite{lorenz}. This model contains the famous Lorenz strange attractor, and exhibits chaotic behavior. It also agrees surprisingly well with three-dimensional (3D) simulations of turbulent convection associated with oxygen burning prior to core collapse \citep{ma07b,am11b}. This approximation does lack multi-mode behavior, as compared with the simulations, which are dominated by five low order modes (see Fig.~1 in \citealt{am11b}); this may affect the accuracy of the representation of intermittency at large scales and of coherent structures.

Our challenge is to simplify this very complex problem, with time dependence and an astronomically large number of degrees of freedom, down to a feasible level for use in a stellar evolutionary code, {\em without losing important features}.
Our approximation, 321D, is an attempt to increase physical realism at feasible cost in computational complexity.
It is desirable to avoid astronomical calibration as far as possible, and base changes upon behavior quantified in laboratory and numerical experiments. In particular, we do not validate our approximation by how well it reproduces standard MLT results. 
By basing approximations on 3D ILES simulations that (1) exhibit turbulence, (2) have non-uniform composition, and (3) resolve dynamic boundary behavior, it is possible to remove some of the vagueness inherent in many theoretical treatments of convection.

We will compare the global properties of turbulent convection from numerical and analytical viewpoints in Section~\ref{sect2}, examine the structure  and nature of boundaries of convection zones in Section~\ref{sect3}, and summarize our conclusions in Section~\ref{summary}. In an appendix we provide a derivation from 3D fluid flow equations for some useful expressions.

\section{Global Behavior of Convection}\label{sect2}

\cite{wda94,ba98,aa00} found that 2D simulations of stellar oxygen burning developed large fluctuations at the boundaries of the convective region. \cite{kwg03} found that 3D simulations of the same stage gave no such boundary fluctuations.
\cite{ma06} did both 2D and 3D simulations and showed that the discrepancy was due to a different choice of boundary condition: \cite{kwg03} used rigid boundaries at the edge of the convective region, while the other simulations included dynamically-active stable layers surrounding the convection, a more realistic choice. Nevertheless, all obtained a convective velocity  of $u \sim 10^7 \rm cm/s$. The global character of the velocity field seemed to be insensitive to the details of the convective boundary, although these fluctuations are an important part of the physics of the boundary itself (and the extent of the convective region). This insensitivity allows us to separate the global problem from the boundary problem (see also \citealt{can92}); in this section we focus on the global problem.

The turbulent kinetic energy equation may be integrated over a convective region;  in the steady state limit this gives a global balance between 
driving on the integral scale, and dissipation at the Kolmogorov scale (see Fig.~\ref{cascade}). This balance has been verified experimentally and numerically as a common feature of turbulence (e.g., \citealt{tennekes,davidson}).
This introduces a length scale, the depth of the convective zone, into the problem.

\subsection{The Turbulent Cascade}\label{cascade_Re}

Using a classical radiative viscosity \citep{mm84}, the Reynolds number is $Re \sim 10^8$ at the base of the solar convection zone\footnote{Using only a classical plasma viscosity due to ion collisions,
the Reynolds number would be even larger \citep{amv14}.}. Numerical simulations and laboratory experiments become turbulent for roughly $Re \geq 10^3$, so fluid flows in stars are strongly turbulent if, as we assume for the moment, rotational and magnetic field effects may be neglected. 

 For homogeneous, isotropic, and steady-state turbulence, the Kolmogorov relation \citep{frisch} between the dissipation rate of turbulent kinetic energy per unit mass $\epsilon_t$, velocity $v$, and length scale $\ell$ is
\begin{equation}
\epsilon_t = {4\over 5} v^3/\ell. \label{kolmog_eps}
\end{equation}
\cite{amy09} found that $\epsilon_t = 0.85\ v_{rms}^3/\ell_{cz}$, where $\ell_{cz}$ is the depth of the convective zone, and $v_{rms}$ is the average convective velocity; see their Eq.~6 and nearby discussion, and references to other studies which report such coefficients. For homogeneous, isotropic turbulence, \cite{kolmg41} predicted a coefficient $4/5$ for a region well away from boundaries.
This factor of 0.8 might change for a strongly stratified region, which would have flow better described by plumes than convective rolls.

Eq.~\ref{kolmog_eps} is a global constraint, averaged over fluctuations, and applies to each length scale $\lambda$ in the turbulent cascade, so
\begin{equation}
\epsilon_t \sim (\Delta v_\lambda)^3/\lambda ,
\end{equation}
for all scales $\lambda$, or,
\begin{equation}
\Delta v_\lambda \sim (\epsilon_t \lambda)^{1 \over 3}.\label{vel}
\end{equation}
so that the velocity variation across a scale $\lambda$ is $\Delta v_\lambda$, which increases as $\lambda^{1 \over 3}$. The largest scales have the largest velocities, and are dominated by
advective transport (macroscopic mixing).

The velocity {\it gradient} across the scale $\lambda$ is
\begin{equation}
\Delta v_\lambda/\lambda \sim \epsilon_t^{1 \over 3}/ \lambda^{2 \over 3},\label{vel_grad}
\end{equation}
and increases with decreasing $\lambda$. The smallest scales have the largest velocity
gradients, and are eventually dominated by microscopic mixing (ionic diffusion, radiative diffusion, and viscosity).
A description of the cascade needs both large and small scales; Eq.~\ref{vel} implies that the
largest (integral) scales have most of the kinetic energy and momentum,
while Eq.~\ref{vel_grad} implies that the smallest scales have the fastest relaxation times, which is consistent with
simulations (e.g., \citealt{amy09}).

\subsection{Limitations of Resolution}
\cite{ll59}, \S32, estimated the number of degrees of freedom in a region of turbulent flow to be $N\sim (Re)^{9/4}$.
Laminar flows with free boundaries become unstable at roughly $Re \sim 10^3$. A direct numerical simulation (DNS) would require well over $10^8$ zones to resolve the cascade for this marginally unstable case. Using $Re \sim 10^{8}$ (see Section~\ref{cascade_Re}), 
implies a need for more than $10^{18}$ zones for the Sun, far beyond current computer capacity. 

There may be a smarter way. Kolmogorov's great insight is that turbulence hides the details of the viscous dissipation by the nonlinear interactions of the cascade, so that the dissipation rate is determined by macroscopic parameters. Simulations show a multimode behavior \citep{ma07b}, but only $N\sim 5$ dominant modes\footnote{See \cite{hlb96} and more recent work on principle component analysis and other techniques which attempt to exploit the reduction in complexity.}
for $\sim 10^8$ zones. This dramatic reduction in complexity suggests the use of implicit large eddy simulations (ILES, see Fig.~\ref{cascade} and \citealt{boris}) which approximate small scale behavior by a Kolmogorov cascade. Our approach is to assume that this simplification holds for very large Reynolds numbers, and to examine the consequences. Simulations which are presently feasible have effective Reynolds numbers limited by numerical resolution, but  are sufficiently high to give truly turbulent solutions.  State of the art simulations, with both improved algorithms and more powerful computers, support this approach \citep{pw00,herwig14,cm15}.

\subsection{Dynamics: MLT to 321D}\label{dynamics}

\begin{deluxetable*}{lllll}
\tablewidth{475pt}
\tablecaption{Correspondence of some variables in MLT, Lorenz and RANS\label{table3}}
\tabletypesize{\small}

\tablehead{ \colhead{quantity$^{a}$} & \colhead{MLT$^{b}$} & \colhead{Lorenz$^{c}$} & \colhead{RANS$^{d}$} 
& \colhead{comment}   
}
\startdata
dissipation length & $ (\alpha^2/8) H_P$ & --- & $ \ell_d\approx 0.8 \ell_{CZ} $ & $\ell_{CZ}$ is convection zone depth \\
 & & & & \cite{ma07b} \\
horizontal$^a$ gradient & $\Delta \nabla=\nabla-\nabla_e$ &$({2 H_P \over \ell}){T_3 / T_0}$ & --- & \cite{sa14} \\
\\
radial gradient & $ \nabla_e-\nabla_a$ &$({2 H_P\over \ell}){T_2 / T_0}$ & --- & \\
\\
imposed gradient & $ \nabla_r-\nabla_a$ &$({2 H_P\over \ell})T_1 / T_0$ & --- & \\
\\
convective velocity & Eq. \ref{mlt} & $u$ & $u'$ & algebraic (MLT) versus ode \\
 & & & & local (MLT) versus nonlocal \\
 turbulent heating & none & ignored or&$\langle (u')^3\rangle/\ell_d $ & \cite{amy09} \\
  & & $u^2|u|/\ell_d$ & &  \\
  \\
kinetic energy flux & assumed & assumed&$\langle \rho' u' {\bf u \cdot u}/2 \rangle$ & \cite{ma07b,ma10}\\
&cancellation&cancellation&  no cancellation,  & \\
&by symmetry & by symmetry &  asymmetry & \\
\\
buoyancy flux &$ u \beta_T g \Delta \nabla$ &${1 \over 2} \beta_T g uT_3/T_0$ & $-g\langle \rho' u'\rangle/\rho_0 $ & MLT ignores composition gradients \\
\\
enthalpy flux & $\rho u C_P T \Delta \nabla$ &${1 \over 2}\rho C_P u T_3$ & $\rho C_P \langle u' T' \rangle$   & \cite{vmam13} \\
\\
acoustic energy flux & none & none & $\langle P' u' \rangle$ & small for low-mach flow \\
\\
composition$^e$ flux & undefined  & none &$\rho \langle Y' u' \rangle $ & \cite{wda96}\\
\\
$Y_e$ flux          & undefined  & none &$\rho \langle Y_e' u' \rangle $ &   \\
\enddata
\tablenotetext{a}{The MLT variables are all defined in the radial direction.  RANS projects a 3D average onto the radial direction. The Lorenz model has both radial and horizontal gradients \citep{am11b,sa14}.}
\tablenotetext{b}{\cite{sa14}.}
\tablenotetext{c}{\cite{amy09,am11b}; $\ell$ is the roll diameter.}
\tablenotetext{d}{\cite{ma07b,vmam13}.}
\tablenotetext{e}{\cite{wda96}, $Y=Y_e+\Sigma_i Y_i=Y_e+1/\bar A$ and $Y_e = \Sigma_i Z_i Y_i$.}

\end{deluxetable*}

As an aid to the reader, Table~\ref{table3} gives the correspondence of selected variables in three different theoretical approaches to turbulent convection: MLT, the Lorenz model, and the RANS formulation. MLT is 1D (radial), the Lorenz model is 2D (radial and transverse), while the RANS analysis is 3D projected to 1D. MLT is static, the Lorenz model and the RANS equations are time dependent. MLT is local (no spatial derivatives of velocity) while the Lorenz model is mildly nonlocal (it uses global derivatives over the roll), and the RANS equations are non-local. Comparison of MLT and Lorenz gives a sense of transverse versus radial properties.

In MLT the buoyant acceleration is approximately integrated over a mixing length $\ell_{MLT}$ to obtain an average velocity $u$ (e.g., \citealt{vitense53,bv58,kippen}), 
\begin{equation}
u^2 = g \beta_T \Delta \nabla  \Big ( {\ell^2_{MLT} \over 8 H_P } \Big ). \label{mlt}
\end{equation}
The superadiabatic excess $\Delta \nabla$ is defined in Table~\ref{table3} and \S\ref{nonuniformY}.
Here $g$ is the gravitational acceleration, $\beta_T = -(\partial \ln  \rho /\partial \ln T)_P$ is a thermodynamic variable  (for uniform composition; see \S\ref{nonuniformY} for the nonuniform case), $H_P$ is the local pressure scale height, and $\ell_{MLT}$ is an adjustable length scale (the mixing length).

Eq.~\ref{mlt} requires that $\Delta \nabla \ge 0$ for the velocity $u$ to be a real number. The velocity depends only on the local value of the superadiabatic gradient $\Delta \nabla$. There are obvious problems with regions in which such integration extends past a boundary.

There have been a number of attempts to generalize MLT; e.g.,  \cite{unno61}, \cite{gough67,gough77}, \cite{wda69}, \cite{stell76}, \cite{kuhfuss86}, \cite{xiong86}, \cite{dbc96}, \cite{xiong97},
\cite{hansen}, \cite{deng06}, etc.
Working backward, Eq.~\ref{mlt} may be expressed as a co-moving acceleration equation for a vector field 
$\bf u$:
\begin{equation}
d{\bf u}/dt =  {\cal B} - {\cal D}, \label{gen_acc}
\end{equation}
where $\cal B$ is a generalized driving term and $\cal D$ a corresponding drag term (\cite{prandtl}, Ch.~V). A
hydrostatic background will be assumed; see Appendix \S\ref{conv_append}.
Similar equations result from (1) study of the nonlinear development of the Rayleigh-Taylor instability (RTI), and from (2) applications of Reynolds-Averaged Navier-Stokes (RANS) analysis to 3D simulations of turbulent convection.

If the driving is due to buoyancy alone,  
(see \S\ref{nonuniformY} for nonuniform composition),
$-g(\delta \rho/\rho) \approx{\bf g} \beta_T \Delta \nabla $,
then
${\cal B} \approx  {\bf g} \beta_T \Delta \nabla$. If the drag is represented by 
${\cal D} \approx {\bf u}/\tau$, where $\tau = \ell_d/|u|$, then
we have
\begin{equation}
d{\bf u}/dt = \partial {\bf u}/\partial t +{\bf (u \cdot \nabla) u} = {\bf g} \beta_T \Delta \nabla - {\bf u}/\tau \label{mlt_acc}.
\end{equation}
This is basically a statement of Newtonian mechanics, with driving by buoyancy and damping by drag. \cite{gough77} gives a historical context going back to \cite{prandtl25} and to \cite{biermann32}. The early attempts, and many of the recent ones, have used a kinetic theory model, in which the mixing length was a sort of mean free path. In contrast, we interpret Eq.~\ref{gen_acc} as a model of the momentum equation for fluid dynamics, involving structures such as waves, convective rolls, or plumes. Because it is non-local, Eq.~\ref{mlt_acc} allows formally stable regions to be convective, unlike MLT, because of finite velocities. This may be relevant for composition mixing in weakly stable regions, and the mass contained in convective regions.

Taking the dot product of Eq.~\ref{mlt_acc} with $\bf u$ gives a kinetic energy equation,
\begin{equation}
d(u^2/2)/dt = {\bf u \cdot g} \beta_T \Delta \nabla - u^2/\tau, \label{mlt_KE}
\end{equation}
for which the steady-state solution\footnote{Care must be taken (for negative $u$) with the sign of the transit time $\tau$ and the deceleration.} is Eq.~\ref{mlt}, with
$\ell_d =  {\ell^2_{MLT} / 8 H_P }$,  and $\Delta \nabla >0$. 
 In Eq.~\ref{mlt_KE}, negative values of $\Delta \nabla$ are allowed; this permits buoyant deceleration \citep{bt02}. The singularities in MLT at the convective zone boundaries (\S9 in \citealt{gough77}), and in boundary layers (\S40 in \citealt{ll59})    are removed\footnote{The singularities in this case occur in Prandtl's equations for a boundary layer as the velocity perpendicular to the surface goes to zero. In a star the motion does not go to zero but becomes wave-like rather than turbulent.}.

The flow is relative to the grid of the background stellar evolution model, so 
the co-moving time derivative of turbulent kinetic energy leads to
\begin{equation}
d(u^2/2)/dt = \partial_t ({\bf u \cdot u})/2 + {\bf \nabla \cdot F_{K} },\label{FKE}
\end{equation}
where $ {\bf F_{K} } = \rho {\bf u} ({\bf u \cdot u})/2$ is a flux of kinetic energy. The generation of the divergence of a kinetic energy flux in this way is robust for dynamic models; it occurs in the more precise RANS approach (Eq.~\ref{TKE} as well as Eq.~\ref{mlt_KE}). 

 We may write Eq.~\ref{gen_acc} as
\begin{equation}
 \partial_t ({\bf u \cdot u})/2 + {\bf \nabla \cdot F_{K} } = {\bf u \cdot ({\cal B - D}) } .
\end{equation}
In a steady state, the divergence of turbulent kinetic energy flux is zero only if there is a {\em local balance between the driving and the drag terms}. Otherwise turbulent kinetic energy flux may be non-negligible.
The turbulent kinetic energy flux smooths the distribution of turbulent kinetic energy between regions in which it is generated in excess, and the whole turbulent region. The drag term is usually relatively smooth in comparison to the driving term, which can be strongly peaked. Turbulent kinetic energy transport is especially important if convection is driven by cooling near the photosphere, so that the (negative) buoyancy is localized and the stratification is strong. \cite{ma10} have shown that stratification enhances the asymmetry in convective kinetic energy flux for driving from the top, and  reduces it for driving at the bottom; see also \cite{sn89,catt91}. 
This asymmetry is small for shallow convective zones, growing with stratification.

This behavior does not occur in MLT, which enforces an {\em exact} symmetry between up-flows and down-flows so that $\bf \nabla \cdot F_{K} = 0$. 
Although simulations of 3D atmospheres exhibit strong downward (negative) net fluxes of kinetic energy, such information was not included in MLT fits for such atmospheres  \citep{trampe07,ts11,magic_MLT_14}. 
Simulations of 3D red-giant atmospheres by \cite{lk12} indicate that the fits to MLT require at least a two parameter family, as have simulations of deeper convection.
In the red giant model in \cite{vmam13}, the downward directed kinetic energy flux reaches 35\%
of the maximum enthalpy flux. 
\cite{sn98} find that their solar model has a downward directed kinetic energy flux which is 10\% of the enthalpy flux. This downward kinetic energy flux must be compensated for by a larger (outward) enthalpy flux. This kinetic energy flux is accompanied by a momentum flux, which affects the convective boundary, as shown in \S\ref{braking}.
These are nontrivial differences relative to MLT, and may have implications which are detectable with asteroseismology as deviations from the predictions of MLT models.

At present, stellar evolution theory has no turbulent heating term. This is inconsistent\footnote{Alternatively one might take the view that this is included in the MLT ``convective flux" by construction, but this conflates different physical effects.} with Kolmogorov theory, which states that turbulent kinetic energy is fed back into  the thermal bath at the rate given by Eq.~\ref{kolmog_eps}. From the viewpoint of a dynamic model (e.g., Eq.~\ref{gen_acc}), this is a ``frictional'' cost of moving energy by convection. \cite{amy09} show that energetic self-consistency requires that the usual stellar evolution equations must be modified to include such a heating term, or equivalently, to explicitly include terms for heating by buoyancy work and divergence of kinetic energy and acoustic fluxes (see \citealt{amy09}, Eq.\,20-22; \citealt{mocak14}, \S21.5, \S21.6). The Kolmogorov term appears as heating in the internal energy equation and cooling (damping) in the turbulent kinetic energy (acceleration) equation. Total energy is conserved; turbulent kinetic energy is transformed into heat.

It may be more convenient to apply the heating term directly, rather than use the buoyancy work and divergence of turbulent kinetic energy and acoustic fluxes, as the velocity is available from solution of Eq.~\ref{mlt_acc}. 
Turbulent heating (and divergence of kinetic energy flux) may have implications for the standard solar model and solar abundances\footnote{\cite{wda14} suggested that the flux of turbulent kinetic energy was simply responsible for a change in radiative luminosity in the solar model. The situation is more complex. The finite negative luminosity of turbulent kinetic energy flow is compensated by an increased positive enthalpy flux, and a radiative flux. This modifies the thermal structure. The turbulent momentum flux in the braking region (\S\ref{braking}) extends the well-mixed region beyond the conventional Schwarzschild estimate; these effects would modify the solar model in the same sense.}. Such heating may also be important for the motion of convective burning shells into electron-degenerate fuel.

In the {\em local, steady-state,} limiting case, the left-hand side of Eq.~\ref{mlt_KE} vanishes, and an equation similar to Eq.~\ref{mlt} results, but with a turbulent damping length instead of a mixing length. In simulations this is the lesser of the depth of the convective zone or $4$ pressure scale heights\footnote{This upper limit to the turbulent damping length may be related to increasing stratification. The development of plumes and their Rayleigh-Taylor instability will enhance the turbulent drag, reducing the increase in $\ell_d$; see \S\ref{rti} } \citep{am11b}.
With this change, {\em the cubic equation of B\"ohm-Vitense may be derived} \citep{sa14},
and we recover a form of MLT.

Had it been available, B\"ohm-Vitense might have identified the mixing length with the Kolmogorov damping length (Eq.~\ref{kolmog_eps}). 
However, Kolmogorov found the damping length $\ell_d$ to be the depth of the turbulent region, so that it is not a free parameter, unlike MLT. There is a further issue: $\epsilon_t$ is the {\em average} dissipation rate, not the instantaneous local value ($u^3/\ell_d$) which fluctuates over  time and space (see Fig.~4 in \citealt{ma07b}); that is, $ u \neq v$ except on average. This is reminiscent of the RANS approach (\S\ref{fluctuations} and \S\ref{TKE_subsection}).

Suppose we assume that the integral scale motion is that of a 2D convective roll,
where $d{\bf u}/dt$ is given by Eq.~\ref{mlt_acc}.  Using this and
a corresponding thermal energy equation, we obtain a form of the classic Lorenz equations, but with a nonlinear damping term provided by the Kolmogorov cascade \citep{am11b}. {\em Because of the time lag, as implied by the time needed to traverse the cascade from integral to Kolmogorov scales, the modified equations are even more unstable than the original ones, and have chaotic behavior}.\footnote{Direct integration shows that, even for no time lag in dissipation, chaos sets in slowly at a Reynolds number Re between 600 and 700.}

\subsection{Nonuniform Composition}\label{nonuniformY}
In Eq.~\ref{mlt_acc} it was assumed that the density fluctuation which drives the buoyancy could be represented by  $-g(\delta \rho/\rho) \approx{\bf g} \beta_T \Delta \nabla $, involving only a fluctuation in temperature.  This is only true for uniform composition and mild stratification. The formulation makes use of the expansion of pressure fluctuation,
\begin{eqnarray}
P' = (\partial P/\partial T)_{\rho,Y} T' + (\partial P/\partial \rho)_{T,Y} \rho' \nonumber \\
 + (\partial P/\partial Y)_{T,\rho} Y',
\end{eqnarray}
which may  be written as
\begin{eqnarray}
\rho'/\rho = -\beta_T (T'/T) -\beta_Y (Y'/Y) \nonumber \\
+ (P/\rho s^2) P'/P,
\end{eqnarray}
where 
\begin{eqnarray}
\beta_T = -(\partial \ln \rho/\partial \ln T)_{Y,P},\\  
\beta_Y = -(\partial \ln \rho/\partial \ln Y)_{T,P},\\
s^2 =(\partial P/\partial \rho)_{T,Y}.
\end{eqnarray}
Here $s$ is the sound speed. The composition variable $Y$ denotes the number of free particles per baryon \citep{wda96}, and is essentially the inverse of the mean molecular weight $\mu$ \citep{kippen,hansen}. An illustrative and simple example is the ideal gas, $P={\cal R} \rho Y T$. For subsonic flows, $|P'/P| \sim (u/s)^2$, where $u/s$ is the Mach number of the flow, and is small\footnote{Near boundaries the approximation $P'/P \sim 0$ fails because pressure fluctuations provide the transverse acceleration necessary to divert the flow; see \S\ref{braking}.}. In MLT, the pressure fluctuation is assumed zero (no acceleration by pressure dilatation), so
\begin{equation}
 \beta_T (T'/T)  + \beta_Y(Y'/Y) \approx - (\rho'/\rho),\label{TYrho}
\end{equation}
and it is further assumed that $Y'=0$ to obtain Eq.~\ref{mlt}.
Even in the limit of negligible pressure fluctuations,
{\em variations in $Y$ enter in a way similar to variations in $T$}, so even small composition variations can be significant when superadiabatic temperature variations are also small. Many of the difficulties found using MLT are related to situations in which $Y'\neq0$: overshooting, semi-convection, and entrainment.

\subsection{Dynamics: Rayleigh-Taylor instabilities}\label{rti}

There seems to be a deep connection between Eq.~\ref{mlt_acc}, Rayleigh-Taylor instabilities (RTI), and turbulent mixing.  An almost identical equation (Eq.~4.1 in
\citealt{abarzhi2010}) is used to  describe the nonlinear development of the RTI into the turbulent mixing regime. Unlike canonical Kolmogorv turbulence, the RT turbulent mixing is statistically unsteady, and involves the transport of potential and kinetic energies as well as enthalpy. Because of its importance in a variety of high energy-density (HED) conditions \citep{zeldovich02,kane97,radt99,dimonte04,remington06,drake09,kuranz11},
much experimental effort for its study as well as an extensive literature have developed.

The RTI happens when a heavier fluid overlays a lighter one, proceeding from linear instability of perturbations \citep{ch61}, to mildly nonlinear motion of bubbles and spikes, and then to nonlinear turbulent mixing \citep{abarzhi2010}. The initial acceleration is one-dimensional, but as instability develops, the motion breaks symmetry and approaches isotropy (as seen in a co-moving frame), much like the cascade in steady turbulence \citep{frisch}.
The essential difference between stellar convection and RTI is that the RTI is not contained, while convection operates within a definite and slowly varying volume. This means that the vertical and the transverse scales are causally connected in convection, but may be independent in the RTI \citep{abarzhi2010}.

Inconsistency between experimental and numerical investigation of the RTI in the nonlinear regime led to the 
$\alpha_b$  problem \citep{dimonte04}. The RTI in the limit of strong mode-coupling can be initiated to have self-similar evolution, so that the amplitude (diameter of the bubble $D_b \propto h_b$) evolves as $h_b \sim \alpha_b A g t^2$, where A is the Atwood number (density ratio, \citealt{ch61}), $g$ is gravity and $t$ the elapsed time. The simulation value $\alpha_b \sim 0.025 \pm 0.003$ is smaller than the experimental value $\alpha_b \sim 0.057 \pm 0.008$.
This discrepancy seems to have been resolved by the idea that unquantified errors in the experimental initial conditions were the cause. To the extent that such uncertainties cannot be precisely known, this suggests a statistical approach, and illustrates the need for combined theoretical, experimental, and numerical studies.  

\cite{ma07b} found that regions of their simulated convection zone exhibited recurring ``bursts" of convection (see their Fig.~4). These bursts, although multi-modal ($n \sim 5$), seem to share the chaotic behavior of the \cite{lorenz} model of a single-mode convective roll \citep{am11b}.
    This encourages the use of Eq.~\ref{mlt_acc}, which is related to the momentum-driven model of RTI \citep{abarzhi2010}, for timescales less than or of order of the transit time. For longer, evolutionary timescales (stellar convection) we need to average over fluctuations, which means averaging over several transit times for the convective roll (see Eq.~\ref{TKE} below). These bursts result from underlying physics similar to that in the RTI; their short timescale behavior may be relevant for stellar pulsations and eruptions (the $\tau$-mechanism, \citealt{am11b}, or equivalently, stochastic excitation of oscillations, \citealt{gk90,gmk94,aerts}).

\subsection{Filtering Fluctuations}\label{fluctuations}

The weak coupling between driving at the large scale, and dissipation at the small scale, allows 
time dependent fluctuations of significant amplitude in luminosity and turbulent velocity.
The term $ \partial {\bf u }/ \partial t$ (Eq.~\ref{gen_acc} and Eq.~\ref{steady_u}) 
is needed for chaotic fluctuations and wave generation. 
These fluctuations have
a cellular structure in space and time; if there are many cells, with random phases, the fluctuations in the average total luminosity are reduced by cancellation \citep{am11b}.

Fluctuations are fundamental features of  turbulence and mixing.  
 Because of sensitivity to initial conditions which can never be known with complete accuracy, descriptions of turbulence should be statistical in nature, even though the equations are deterministic \citep{frisch}. Turbulent simulations can be said to be numerically converged only in a statistical sense. Eventually trajectories will diverge.
 Lyapanov exponents characterize this divergence, a feature characteristic of turbulence \citep{manneville} which makes turbulent mixing so effective.
 Unlike the diffusion picture, in which a stellar mixing front moves radially, limited by the random walk of mean-free-path strides, turbulent mixing involves a network of trajectories throughout the space of the turbulent region, laced with inhomogeneities, which finally disappear at the Kolmogorov scale.

\begin{figure}[h]
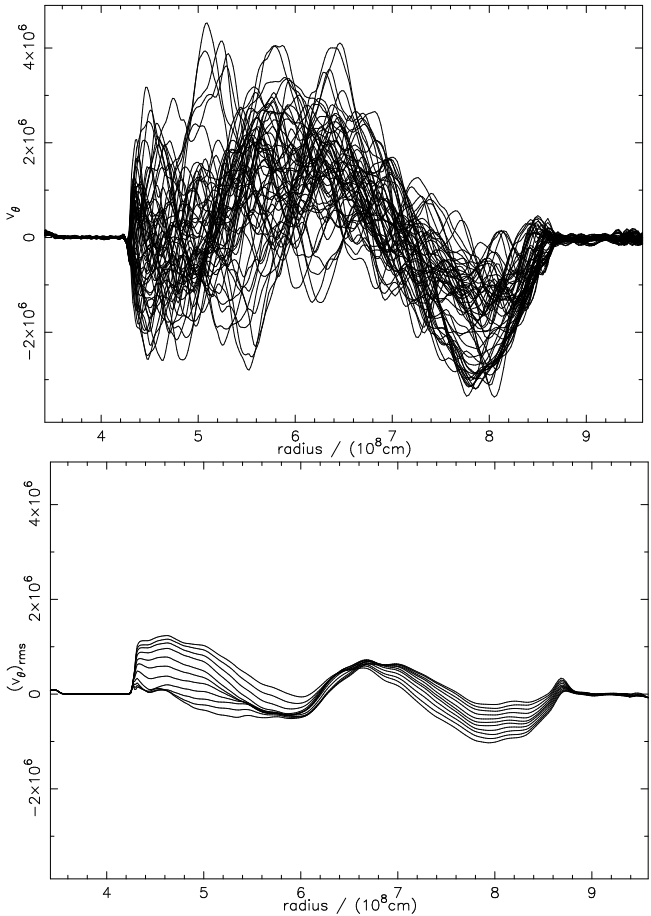

\figurenum{2}
\includegraphics*[angle=-90,scale=0.35]{figure2a.ps}
\includegraphics*[angle=-90,scale=0.35]{figure2b.ps}
\caption{ Fluctuations in velocity $v_\theta$ (in cm/s) versus radius. The top panel shows a sequence of snapshots, one for each time-step of $\delta t \sim 0.5 \rm s$. The lower panel shows the running average over 300 such steps (150 s), starting at 12 times, each separated by 20 steps (10 s). The vertical scales are identical. Averaging gives linear cancellation of small time scale fluctuations, but longer time scale variations survive. See text and \cite{vmam13}, model OB. 
The simulation shows a flow involving several ($\sim 5$) prominent modes which decay and reform. 
See text for details.
}
\label{fig_fluct}
\end{figure}
\placefigure{2}

In stratified regions, mass conservation constrains the flow, but it tends to change the cross-sectional area of the plumes as opposed to limiting their range. Although the flow is locally wild with fluctuations, these tend to cancel upon horizontal and time averaging, leaving a much more placid behavior due to the cancellation of random phases.
Fig.~\ref{fig_fluct} illustrates this for a particular but representative case; the velocity
in the theta direction, $v_\theta$, is shown as a function of radius, from the oxygen burning data set in \cite{vmam13}. The top panel shows the instantaneous value of $v_\theta$ (in units of cm/s) for a sequence of time steps $\delta t \sim 0.5 s$. The bottom panel shows the running average (a horizontal average, i.e., over a spherical surface of radius $r$) 
of the same variable over 300 such time steps (150s), stepping forward over 20 time steps (10s) at a stride, on the same velocity scale. The amplitude in the bottom panel is much reduced by cancellation; what does remain is the larger length scale, as suggested by the cascade idea discussed in \S\ref{cascade_Re}. The cancellation does not work for quadratic terms; they remain non-zero, e.g., contributing to the rms velocity in this case (see \S\ref{TKE_subsection}). The product of fluctuations in velocity and temperature give rise to the enthalpy flux; those  in velocity and composition give rise to the composition flux.

A stellar evolution code must step over the shorter turnover time scales (weather) to solve for the evolutionary times (climate). How can this be done?
 It requires an average over active and inactive cells.
 The steady-state limit of the Lorenz equation seems to give a reasonable approximation to its average behavior, filtering out the chaotic fluctuations \citep{am11b}.
Instead of $d{\bf u}/dt=0$, we use
\begin{eqnarray}
d{\bf u}/dt &= {\bf \partial u /\partial {\it t}} +  ({\bf u \cdot \nabla) u  },\label{partial_u} \nonumber \\
 & \rightarrow {\bf (u \cdot \nabla) u  }.\label{steady_u}
\end{eqnarray}
We apply the same approximation (Eq.~\ref{steady_u}) to Eq.~\ref{mlt_acc} for slow stages of stellar evolution.
This allows non-local behavior, will prove important for our discussion of convective boundaries later in \S\ref{sect3},
and can represent ram pressure (Reynolds stress) and the flux of turbulent kinetic energy; see also \S3.2 in \cite{pw00}, for
a discussion of ram pressure in 3D simulations relative to MLT.
 
Now we have established connections between an acceleration equation (Eq.~\ref{gen_acc}) and (1) MLT, (2) historical attempts to extend MLT,  (3) modern research on RTI \citep{abarzhi2010}, (4) the important advances of \cite{kolmg41,kolmg} and \cite{lorenz}, and (5) a rational way to step over fluctuations for stellar evolution.

\subsection{Turbulent Kinetic Energy Equation}\label{TKE_subsection}

A more rigorous alternative is to use the Reynolds-averaged Navier-Stokes (RANS) approach, which directly averages the fluctuations over space and time. This has been explored by Canuto \citep{can11a,can11b,can11c,can11d,can11e}, see also \cite{xiong97,deng06}; a detailed comparison with their work, while desirable, is beyond the scope of this paper. Canuto uses simulations and experiments from geophysics to effect a closure of the RANS equations, while in contrast, our closure of the RANS is based on our 3D simulations. 

The turbulent kinetic energy equation (TKE) is obtained by a Reynolds decomposition of the velocity, density, and pressure (detailed discussion may be found in \citealt{ma07b,amy09,vmam13,mocak14}). 
In principle the TKE is exact; errors arise from closure, i.e., our 
analytical approximations to the terms in the RANS equations are at fault. 
Well-resolved 3D ILES simulations show excellent agreement with the TKE \citep{vmam13}, and allow the dominant terms to be identified. Being more general than the simpler approximations discussed above, the TKE allows us to identify and quantify neglected terms. Most importantly, it allows an enormous simplification and compaction of the 3D numerical data, while that data in turn allows a closure of the RANS procedure.

The TKE may be written as \citep{ma07b}:
\begin{eqnarray}
\partial_t \langle\overline{ \rho E_K}\rangle  + {\bf \nabla \cdot}\langle\overline{\rho E_K {\bf u_0}}\rangle 
= \nonumber\\
-{\bf \nabla \cdot} \langle\overline{\bf F_P + F_K}\rangle +\langle\overline{P' {\bf \nabla \cdot u'}}
\rangle \nonumber\\ 
+ \langle\overline{\rho'{\bf g \cdot u'}}\rangle - \rho \epsilon_d, \label{TKE}
\end{eqnarray}
We use $\langle q\rangle$ and $\overline{q}$ to denote angular and time averages of a quantity $q$. Primes refer to fluctuating quantities; for example $\bf u = u_0 + u'$, and $\langle \bf u \rangle = u_0$, and similarly for the time average. The turbulent  kinetic energy per unit mass is $E_K = {1 \over 2}( {\bf u' \cdot u'} )$, a measure of the rms turbulent velocity. The acoustic and turbulent kinetic fluxes are ${\bf F_P} = P'{\bf u'}$ and ${\bf F_K} = \rho E_K {\bf u'}$. The dissipation may be written as 
\begin{equation}
\epsilon_d = \langle \overline{ {\bf u' \cdot u'} |u'| } \rangle / \ell, \label{damping}
\end{equation} 
a form which we identify with Eq.~\ref{kolmog_eps}, the expression of \cite{kolmg41,kolmg}; notice that it involves averages of powers of the velocity fluctuation, not the instantaneous values.

Using the RANS approach is equivalent to using the bottom panel in Fig.~\ref{fig_fluct} rather than the top; it removes the fluctuating activity which cancels (has no net effect), while keeping what does not cancel.

To better understand the implications of the TKE, consider (1) a steady state ($\partial_t \langle\overline{ \rho E_K}\rangle = 0$) with (2) no background motion ($\bf u_0=0$). Then the TKE reduces to the divergence of the fluxes ${\bf \nabla \cdot} \langle\overline{\bf F_P + F_K}\rangle,$ balancing the net result of two source terms $\langle\overline{P' {\bf \nabla \cdot u'}}
\rangle $ and $ \langle\overline{\rho'{\bf g \cdot u'}}\rangle$, and a damping term  $- \rho \epsilon_d$:
\begin{eqnarray}
{\bf \nabla \cdot} \langle\overline{\bf F_P + F_K}\rangle =\langle\overline{P' {\bf \nabla \cdot u'}}
\rangle \nonumber\\ 
+ \langle\overline{\rho'{\bf g \cdot u'}}\rangle - \rho \epsilon_d. \label{ssTKE}
\end{eqnarray}
This may be integrated over the convection zone (taking the surface fluxes to be zero or small at the boundaries), and if we ignore the pressure dilatation $\langle\overline{P' {\bf \nabla \cdot u'}}$
 for the moment, gives an expression for the damping length $\ell_d$,
\begin{equation}
\ell_d =
 \int_{CZ}\langle\overline{( {\bf u' \cdot u'})^{3 \over 2}}  \rangle dm \Big / \Big (
\int_{CZ} \langle\overline{{\rho'\over \rho_0}{\bf g \cdot u'}}\rangle \Big) dm, \label{ell}
\end{equation}
which is a global condition that must be satisfied to be consistent with Kolmogorov damping, which also requires that $\ell_d$ is approximately the depth of the turbulent region. This characteristic length scale is a fundamental property of turbulence, and is generated robustly in the numerical simulations.

Eq.~\ref{ell} might be regarded as a generalization of the \cite{roxburgh,roxburgh92} integral constraint to include  damping by turbulence.
Notice that $\ell_d$, which appears in both Eq.~\ref{mlt_acc} and Eq.~\ref{ell}, must be solved for consistently; it tends to be a slowly-varying function, of order of the convective zone depth.
Eq.~\ref{ell} involves some of the important ``bulk'' properties discussed by \cite{can92}, and is a statement of a global balance between driving and damping.

What approximations would be necessary to make the TKE equation equivalent to MLT?
In MLT, (1) the net flux of turbulent kinetic energy $\bf F_K$ is defined to be zero by symmetry, (2) pressure fluctuations are ignored so the acoustic flux $\bf F_P$ and pressure dilatation 
$ \langle\overline{P' {\bf \nabla \cdot u'}} \rangle $ are zero, and (3) the damping length $\ell_d$ is taken to be an arbitrary adjustable parameter. Enforcing these gives
\begin{eqnarray}
 \langle\overline{\rho'{\bf g \cdot u'}}\rangle = \rho \epsilon_d, 
\end{eqnarray}
This is the local version of the global balance in Eq.~\ref{ell};
it is equivalent to the B\"ohm-Vitense cubic equation of MLT for the appropriate choice of mixing length \citep{sa14}.

This approximation leads to a series of errors:  (1) Symmetry between up-flows and down-flows is broken by stratification, so that turbulent kinetic energy fluxes are not generally zero \citep{sn89,catt91,can92}. This is a {\it qualitative} error.
(2) Pressure fluctuations may not be ignored for strongly stratified convection zones.
This is a {\it quantitative} error.
\cite{vmam13} find that acceleration by the pressure  dilatation
term is comparable to that from buoyancy. (3) The damping length may not be freely adjusted if the relation of \cite{kolmg41,kolmg} is to be satisfied. Such adjustments are usually necessary to compensate for a lack of non-locality 
in atmospheres due to the lack of ram pressure, and deeper into interiors due to a lack of kinetic energy flux (the two parameters discussed in regard to 3D atmospheres in \S\ref{dynamics}).

\subsection{The \cite{pasetto14} model}\label{pasetto}
Our efforts have been three-fold: (1) construction of accurate numerical solutions of the Navier-Stokes equations which exhibit turbulence, (2) theoretical analysis of these solutions in the RANS framework to determine the most important features, and (3) invention of simpler analytic representations which capture the essential features of the numerical solutions.
\cite{pasetto14} have presented a novel analytical theory of convection in stars which does not contain a mixing-length parameter; this is an alternative to (3) above, and it is of interest to compare how well it agrees with both our numerical solutions (1 and 2), and our analytic approximations (3). 

As we have shown in  \S\ref{TKE_subsection}, the natural length scale for convection is the dissipation length for the turbulent cascade. Part of the foundation of the model of \cite{pasetto14} is the use of potential flow and the Bernoulli equation (\cite{ll59}, Eq.~10.7 in \S{10}), which result from the Euler equation, not the Navier-Stokes equation. Their theory seems to be equivalent to assuming the process occurs on a scale much less than the size of the convective region, so that there is no way to define a length scale for turbulent dissipation. In contrast, following Kolmogorov (\S\ref{cascade_Re}),  the length scale in our theory is the size of the turbulent region, which is not arbitrary but determined by the turbulent flow. Our length scale is not an assumption (as in MLT) but a consistent and robust result of our simulations. It is the length scale over which driving and damping of turbulence balance 
(\S\ref{dynamics}). In order to describe the turbulent cascade, a complete theory must  deal with the whole turbulent region. 

Is the theory of \cite{pasetto14} 
physically correct?  Stellar interior convection is extremely turbulent, so the question becomes: what are the errors introduced by ignoring turbulence?  \cite{ll59} give a careful discussion of the applicability of potential flow (their \S{9}), and they note that the validity of Bernoulli's equation is limited because of the formation of boundary layers in which viscous effects must be included (see also \citealt{prandtl}). Stars have large Reynolds numbers, so  that turbulent boundary layers form (\citealt{ll59}, Chap.~III), as they do in our simulations (Fig.~\ref{qvsr}). The Pasetto theory, like MLT, ignores boundary layers and turbulence, as well as composition gradients.

A basic assumption of the \cite{pasetto14} theory is that velocities of lateral expansion are much larger than those of the vertical rise of convective elements (their \S4.2). However, the simulations show average velocities in the turbulent region which are not strongly biased toward the laterial directions; this was already clear in \cite{ma07b}, (their Fig.~6), and has held true for subsequent simulations with refined resolution \citep{vmam13,cm15}. The rms velocity in the radial direction is actually {\em larger} than the lateral rms velocities, rather than smaller \citep{amy09}.
 
 A key test presented in \cite{pasetto14} of their theory is a comparison with MLT\footnote{As our title suggests, we attempt to go beyond MLT.}  at $r=0.98R_\odot$, well inside the super-adiabatic region (SAR) at $r \sim 0.9985R_\odot$ in the Sun. It is the inefficient convection in the SAR which determines the solar radius in calibrations of stellar evolutionary codes, so that a test in the SAR would be instructive. \cite{pasetto14} state ``Convective elements in this region have low thermal capacity, so that the super-adiabatic approximation can no longer be applied, and the temperature gradient of the elements and surrounding medium must be determined separately". The theory in its present form may not yet be applicable to the SAR.
 
 The value of the Pasetto theory may prove to lie in its significant conceptual differences from MLT, and in its use as a null case to provide insight into the effects of turbulence.
 
\section{Boundaries and Boundary Layers}\label{sect3}

It has been assumed that because deep convection is adiabatic, MLT may be used without problem for standard stellar evolution in deep interiors. This ignores the effects of the velocity field.
Realistic boundary physics requires more than the adiabatic assumption; it requires dynamics to define the boundary, and hence the size of the convective regions \citep{wda94,aa00,ma06,ma07b}.

 Because, unlike MLT, Eq.~\ref{mlt_acc} and its variants have a spatial derivative, {\em the edges of the convective zones may be found  by simply integrating the acceleration equation to find the zeros of the velocity. }

In this section we begin by discussing several issues related to boundaries. We stress the importance of P\'eclet number variation (\S\ref{peclet_num}). We critically review current practice regarding artificial diffusion, real diffusion, semi-convection, and imposed boundary criteria (\S\ref{eggleton_diff}, \ref{michaud},
\ref{semiconv}, \ref{imposed_bnd}). Then we discuss the similarities and differences between convection in stellar atmospheres and deep interiors (\S\ref{atmos_conv}). In \S\ref{deep_conv} we present new numerical results concerning convective boundaries (the development of braking regions, which do not appear in MLT). In \S\ref{braking} we then analyze these results, showing that they emerge from simple considerations of physics, which may be used to construct approximations for use in stellar evolutionary codes.

\subsection{P\'eclet number: radiative diffusion\label{peclet_num}}

For the oxygen-burning shell, the temperature $T$ 
has an abrupt jump inside the mixing region (radius $r \sim 4.3 \times 10^8 \rm\ cm$ in Fig.~\ref{qvsr}). Pressure is continuous through the boundary containing this transition, so that the density curve has  a corresponding dip; see Fig.~2 in \cite{ma07b} or Fig.~5 in \cite{vmam13}. This implies a  steep increase in entropy; 
as evolution continues this entropy jump grows, and the transition region narrows. Such steep gradients in $T$ are a consequence of cooling by neutrinos. They are not seen in earlier, photon-cooled stages of evolution 
and can only be supported for times short compared with timescales for thermal diffusion and electron heat conduction. This is easily the case for oxygen burning because of high opacity and short evolutionary times ($\sim 10^5 \  \rm sec$).  

The P\'eclet number is defined as the ratio of the advective transport rate to the diffusive transport rate of the physical quantity being transported, which here we take to be thermal energy, so
$$Pe = \rm {thermal\ advection\ rate \over thermal\ diffusion\ rate}.$$ In oxygen burning,  radiative diffusion is slow while advection occurs rapidly, giving large P\'eclet numbers (formally infinite since radiative diffusion was small enough to be neglected in some simulations; the infinity results from the denominator in the definition being a negligible term, not from any exceptional behavior of the physics). 

This contrasts with the situation in stellar atmospheres, in which the radiative diffusion becomes faster than advective transport, so that $Pe < 1$. This difference in P\'eclet numbers suggests the possibility of a {\it fundamental flaw in the notion that observations of stellar atmospheres may be sufficient to define the nature of deep stellar convection.}  See discussion in \cite{zahn91}; \cite{vmpa15}.

\subsection{Artificial diffusion}\label{eggleton_diff}

Peter Eggleton  took an early step in dealing with steep gradients in composition, with the introduction of a diffusion operator which he stressed was ad-hoc \citep{eggleton}. This numerically advantageous procedure has been widely adopted for stellar evolution, even though it has the potentially worrisome mathematical property that it increases the order of the spatial derivatives in the equations to be solved.
The \cite{eggleton} equation is
\begin{equation}
{d\over dm}\Big ( \sigma {dX \over dm} \Big) = {DX\over Dt} - {\cal R} \label{eggleton}
\end{equation}
where $X$ is the mass fraction, $m$ is the lagrangian mass coordinate, $\sigma = v_{ML} \ell_{ML}(4 \pi r^2 \rho)^2$ is the effective diffusion coefficient, and $\cal R$ is the nuclear reaction network matrix \citep{wda96}. 
This is equivalent to modeling convection as ``turbulent diffusion.''
The left-hand side is the heuristic diffusion operator; the right hand side is the reaction network operator. The actual composition flux is related to the co-moving derivative on the right-hand side; see \citealt{wda96}, \S4.6. Eggleton integrates over the convection zone to eliminate that spatial derivative; usually it is simply ignored in stellar codes.

The Eggleton approach is equivalent to approximating the composition flux
\begin{equation}
F_Y = \rho {\cal A} \overline{\langle u' Y' \rangle } \label{Yflux}
\end{equation}
by a ``down-gradient'' expression (critically discussed by \citealt{can92}),
\begin{equation}
F_Y \rightarrow \rho {\cal A}\ u( -\ell \partial Y/\partial r ). \label{downgrad}
\end{equation}
Direct comparison with simulations shows that this can  be 
qualitatively wrong (by two orders of magnitude). 
For a contact discontinuity (\citealt{ll59}, \S81), $F_Y \rightarrow \rho {\cal A} u \Delta Y$, as in Eq.~\ref{Yflux}, not  $\rho {\cal A} u (-\ell/\Delta r)\Delta Y \rightarrow \infty$, as in Eq.~\ref{downgrad}.
Proper scaling requires that $\ell \rightarrow \Delta r$ at a boundary if Eq.~\ref{downgrad} is used.

As Eggleton intended, the algorithm smooths steep  gradients, but sometimes faster than real physical processes do, as Eggleton warned.  To the extent that gradients in abundance need to be correctly represented (e.g., for ionic diffusion, or density structure), the down-gradient approximation (in Eq.~\ref{eggleton} and Eq.~\ref{downgrad}), is questionable. In particular, fluxes directly computed in simulations \citep{ma07b,vmam13} show that {\em the down-gradient approximation fails in boundary layers} \citep{mocak14}.

\subsection{Ionic diffusion}\label{michaud}
While real atomic (ionic) diffusion is thought to be slow in stars, the diffusion operator is second order in space derivatives, so that it becomes important in steep composition gradients, i.e., boundaries. 
Georges Michaud has led in the application of true  diffusion processes and radiative levitation to stellar evolution \citep{michaud70,michaud91,mrr07}.
Recently these processes have been applied to horizontal branch and sdB stars
 \citep{mrr05,mrr07,Hu2008, Hu2009, Hu2010, Hu2011,mrr11, Bloemen2014}.
Gravitational settling \citep{Hu2009} and radiative 
levitation \citep{Hu2011} are important to (1) recover the iron-group opacity bump that excites the pulsations 
\citep{Charpinet1997} in those stars,  (2) obtain the correct position of the instability strip in the $\log g - T_{\rm{eff}}$ diagram, and (3) help in understanding their observed atmospheric abundances \citep{mrr11}. 

Because the \cite{eggleton} diffusion uses a difference operator similar to that for ionic diffusion (second order in space), and may reduce the gradients which drive that diffusion, care should be taken that the algorithmic diffusion does not cause errors in the real diffusion (e.g., see \citealt{sga15}).

\subsection{Semi-convection}\label{semiconv}

In stellar physics, the idea of semi-convection has spawned various algorithms (e.g., \citealt{sh58,stothers63,cgr71a,cgr71b,demarquem72,sg76,dorman_rood93}), some of which seem to be physically and numerically inconsistent with others. 
The term ``semi-convection'' refers to a mixing process which occurs in a region that is stable according to the Ledoux criterion but unstable according to the Schwarzschild criterion. It generally is thought to involve mixing of composition, but not significant enthalpy. The composition profile may be adjusted to marginal stability according to the Ledoux criterion.

Semi-convection is also often discussed as a double diffusive instability, involving an interaction between radiative diffusion and ionic diffusion \citep{spruit13,lattanzio14}. Although both radiative and ionic diffusion may be included in a 1D stellar code, this does not capture their interaction and 3D dynamics. Semi-convection may be related to oceanic phenomena (thermohaline mixing) in which heat flow and salt concentration play the doubly-diffusive roles, and which have a long history of study (e.g., see Chap.~8 in \citealt{turner73}; \citealt{gill82}). 
\cite{rosenblum11,pascale} give an extensive discussion with numerical simulations based on the oceanic model, and conclude that, while the problem can be solved in the planetary range of parameter space, the stellar case requires a large extrapolation. This difficulty may be further exacerbated by the indication  that many such regions in stars are bathed in a flux of g-mode waves \citep{ma07b}, which are a nonlocal effect that may complicate the analysis in a nontrivial way \citep{mocak14}.

Even with these uncertainties, there are energetic constraints (see Eq.~\ref{E_mix}) which must be obeyed. The amount of mixing possible is limited by the energy available to mix, which is generally taken to be related to the excess $\nabla_r-\nabla_a$, so that luminosity is used to supply the energy required to mix.

\subsection{Imposed Boundaries}\label{imposed_bnd}

MLT, as a local theory, must be supplemented by additional assumptions about behavior at the boundaries of the convection zone \citep{spie71,spie72}. These are usually discussed in terms of {\it linear} stability theory, i.e., in terms of the Ledoux and the Schwarzschild criteria \citep{kippen} being positive.
The Schwarzschild criterion for convective instability is defined by
\begin{equation}
{\cal S} = \nabla_r - \nabla_a. \label{schwarz}
\end{equation}
Here $\nabla_r$ is what the dimensionless temperature gradient would be if all the luminosity were carried by radiative diffusion and $\nabla_a$ is the adiabatic gradient (see Appendix). 
The Ledoux criterion for convective instability has a composition dependence, and
is defined by
\begin{equation}
{\cal L} = \nabla_r - \nabla_a - {\beta_Y \over \beta_T }\nabla_Y. \label{ledoux}
\end{equation}
The last term is written as ${\phi \over \delta}\nabla_\mu$ by \cite{kippen}, \S6.1, their Eq.~6.12. The $\beta$ factors are defined as in \S\ref{nonuniformY} above. Notice that positive ${\beta_Y \over \beta_T }\nabla_Y$ and positive ${\phi \over \delta}\nabla_\mu$ both inhibit mixing.

Neither of these choices seems satisfactory. They have no dependence upon the vigor of the flow on the unstable side of the boundary, which clearly must make a difference. 

Linear perturbation theory examines the instability of a stable region, treating both sides of the boundary equally. In reality they differ: one side is convective. 
The stiffness of the non-convective side is measured by the Brunt-V\"ais\"al\"a (buoyancy) frequency $N$, (see Eq.~6.18 in \citealt{kippen}, and Eq.~3.73 in \citealt{aerts}), where
\begin{equation}
N^2 = - {\beta_T g \over H_P}(\nabla_e - \nabla_a -(\beta_Y/\beta_T)\nabla_Y) =  - {\beta_T g \over H_P}{\cal L}.\label{brunt2}
\end{equation}
$N$ is the frequency of elastic rebound from a perturbation; it is imaginary in convective regions.
Here $\nabla_e$ is the dimensionless temperature gradient relevant\footnote{The exact meaning depends upon the assumed flow, and is different for MLT and the Lorenz model (see \citealt{am11b,sa14}; and Table~\ref{table3}).} to the perturbed element.
On the non-convective side of the boundary, it may be the same as $\nabla_r$ above, giving the second equality, which refers to the tendency to restore stability in the radiative region.

A delicate point is the value of $\nabla_Y$ near the boundary \citep{arlette}. 
By what mechanism does mixing occur? What is the structure of the partially mixed region of transition between well-mixed and unmixed?
 Present practice in stellar evolution is to use the Schwarzschild criterion, which has no $\nabla_Y$, so that these issues may be ignored, or to use the Ledoux criterion with one of the prescriptions for semi-convective mixing (see 
\S\ref{semiconv}).

Such interfacial issues have long been studied in the fluid dynamics and geophysics communities; see \cite{turner73} for an extensive discussion. The Richardson number is defined as some measure of
$$ Ri = {\rm {potential\ energy\ needed\ to\ mix \over kinetic\ energy\ available\ to\ mix }}.$$ 
The linear condition for ability of a layer to resist shear is the
``gradient'' Richardson number $Ri$.
\begin{equation}
Ri = N^2/(\partial u/\partial r)^2  >{1 \over 4} \label{Ri_gradient}
\end{equation} 
is stable; larger stiffness ($N^2$) and less swirling ($(\partial u/\partial r)^2$) tend toward stability. In their discussion of entrainment, \cite{ma07b} used a ``bulk'' (i.e., non-local and non-linear) Richardson number which involved an integral over the region around the boundary. 

In the absence of global rotation, a layer having constant total entropy\footnote{See, e.g., \cite{wda96} for explicit derivations of all components of the entropy (Appendix B), and of the total energy of the star (Appendix C).} is energetically neutral with regard to mixing.
 If after a mixing episode, the luminosity returns to its value for radiative balance ($\nabla_r$ is unchanged), then the additional energy\footnote{This is the change in internal energy due to composition change, keeping temperature and pressure constant.} required to remove the stable compositional stratification is 
\begin{equation}
E_{mix} = g H_P\beta_T(  {\cal L-S}) = g H_P (-\beta_Y \nabla_Y).\label{E_mix}
\end{equation}
Both $\beta_T$ and $\beta_Y$ are intrinsically negative in stars. 
If this energy $E_{mix}$ changes sign,
mixing may occur which is driven by the gradient in composition \citep{mocak11b}.
Using a {\bf specific} kinetic energy of ${1 \over 2}u^2$, a Richardson number may be constructed,
\begin{equation}
Ri = 2 g H_P(- \beta_Y \nabla_Y )/ u^2. \label{Ri_general}
\end{equation}
Here the traditional $Ri > 1/4$ is a plausible condition for stability, at least roughly.

\subsection{Solar  convection}\label{atmos_conv}

In their pioneering work on 
solar convection, \cite{sn89} carefully explored the topology of convective flow below the photosphere: converging, cool downdrafts being dominant, with radiative cooling providing the entropy deficit which drives the circulation.
\cite{fls96} examined shallow (weakly stratified) convection, driven by atmospheric cooling, and emphasized the importance of the atmosphere in determining the nature of the convection zone. 
As deep interior convection \citep{wda94,ba94} has no atmosphere, atmospheric physics can have no strong role there (the circulation is driven by nuclear burning). Furthermore, the bottom boundary, which could be ignored in the simulations of \cite{sn89}, may be important for the detailed effects of solar convection on the interior.

\cite{schw58}, \S11, showed that, for stellar interior models, the atmosphere could be represented by an entropy jump between the photosphere and the adiabatic (deep) convective region.  This entropy jump is a primary parameter for determining the depth of the convection zone.  
The atmospheric model 
is crucial for predicting spectral features for a given entropy jump, but has a weak influence on that entropy jump itself \citep{tanner12,tanner14}. 

Many features of the atmospheric and deep interior simulations are similar, leading to the idea that atmospheric physics, however crucial for spectral formation \citep{sn98,magic_I_13,magic_MLT_14}, may be treated as a boundary condition issue rather than a key feature of deep turbulent convection.  \cite{ma10} showed that the general characteristics of the flow in solar convection  (narrow, fast down-flows with broad, slow up-flows and acceleration by pressure dilatation, \citealt{sn89,vmam13}), require only localized top cooling and stratification. 
Global simulations of the solar convection zone are necessarily less well resolved for comparable computational resources; the simulations of \cite{mbdt07} are beginning to show turbulence, but may require finer zoning to deal with some details of the turbulent flow (e.g., \citealt{hanasoge,brand15}).

\subsection{Deep interior convection}\label{deep_conv}

\begin{deluxetable*}{lllll}
\tablewidth{420pt}
\tablecaption{ Integral Properties of Convection Zone Regions\label{table2}}
\tabletypesize{\small}

\tablehead{ \colhead{variable} & \colhead{symbol} &\colhead{total (CZ+BL+BL)} & \colhead{lower BL} 
& \colhead{upper BL}   
}
\startdata
mass & $\Delta m/M_\odot$ & 0.9205& 0.0161 & 0.1150 \\
depth &$\Delta r/10^8$cm     & 4.460 &0.078 & 0.587 \\ 
kinetic energy &KE/$10^{46}$erg & 8.608 & 0.255 & 0.561 \\
buoyancy luminosity &$L_{buoy}/10^{45}$erg/s& 4.576 & -0.0342 & -0.0492 \\
pressure &$ \Delta \ln P$ & 2.032 & 0.046 & 0.228 \\
number of zones$^{a}$& $\Delta i$ & 236 & 8 & 23 \\
\enddata
\tablenotetext{a}{The total number of zones in the radial direction was 400 in \citealt{ma07b} (this table, medium resolution),  800 in \citealt{vmam13} (high resolution), and 1536 in Fig.~\ref{uhi-res_qvsr}. The basic features
appear even at lower resolutions.}

\end{deluxetable*}

\placetable{3}

The simplest of stellar convection zones are cooled by the local processes  (cooling by neutrino emission and heating by nuclear burning), rather than the non-local processes (radiative transfer), giving a cleaner example of the dynamics of boundaries for deep convection. A slightly more complex case is a convection zone with heat conduction by radiative diffusion; \cite{vmam13} consider both. These two cases cover almost all of the conditions relevant to stellar evolution, except the outer layers simulated in 3D atmospheres.

For the oxygen-burning shell, some integral properties of the main convective region and the braking layers
are summarized in Table~\ref{table2}.
About 14 percent of the mass and 15 percent of the thickness of the total convection zone are in the boundary layers (upper BL and lower BL), as is 8.5 percent of the turbulent kinetic energy.
These boundary regions provide deceleration (braking) of the vertically directed flow, allowing it
to remain bounded by the convective volume. If the buoyancy flux is
$q =  -g \langle u'_z \rho'\rangle  /\rho_0 $, 
then the rate at which turbulent kinetic energy increases due to buoyancy in a region $a$, is 
\begin{equation}
L_{buoy}(a) = \int_{a} q  \ dm,
\end{equation}
 which is positive in the middle region, but negative in the boundary regions. These regions of negative buoyancy are a robust qualitative feature of the simulations, dating back to early 2D work \citep{hurl84,wda94}. In the oxygen-burning shell they reduce the driving of turbulent kinetic energy by only 1.8 percent. 
  
Table~\ref{table2} shows the depth of each region in pressure scale heights ($\Delta \ln P$).
The depth of the boundary zones is not a universal constant in $\ln P$, but varies by a factor of ~5 between top and bottom. The last line gives the number of zones in each region for ``medium'' resolution \citep{ma07b}; the lower boundary region is most demanding, having a steep transition from convective to stable stratification. 

Little of the kinetic energy 
is lost in the boundary regions, so $L_{buoy}$ provides a good first estimate of the rate of generation of turbulent kinetic energy.  These regions contain $17\%$ of the mass in the ``convection zone"; most of this comes from the upper layer, which has less extreme stratification.

\begin{figure}[h]
\figurenum{3}
\includegraphics[angle=-90,scale=0.35]{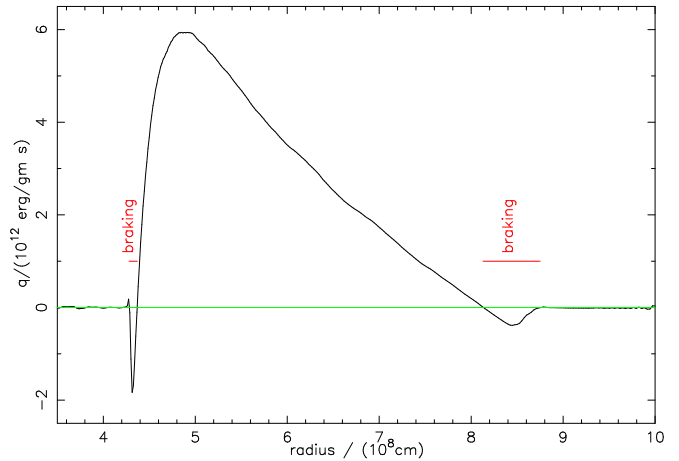}

\caption{Buoyancy Braking averaged over 100 seconds ($\sim 2\rm\, transit\, times$) at shell boundaries for oxygen burning: $q$ versus radius \citep{ma07b,vmam13}. The buoyant acceleration changes sign near the boundaries of the convection zone, giving braking rather than positive acceleration. 
}
\label{qvsr}
\end{figure}
\placefigure{3}

Fig.~\ref{qvsr} shows the buoyancy flux
versus radius, averaged over 100 seconds, for the oxygen-burning shell simulation (OB); more detail may be found in \citep{ma07b,amy09,vmam13}. 
The buoyancy flux, $-{\bf u \cdot g } \rho' /\rho_0 ,$ is the rate of work done by gravity \citep{zahn91}. It is
 the rate of flow of buoyancy, $-{\bf g } \rho' /\rho_0 ,$ and has units of energy per unit mass per unit time (e.g., erg/g/s). Over most of the convective region it is proportional to the enthalpy flux \citep{amy09}.

Fig.~\ref{qvsr} shows that the convective zone simulation is naturally split into three regions, separated by two boundaries.  The regions above and below are stable.
The middle region is relatively uninfluenced by the boundaries; it
is characterized by positive fluxes of buoyancy and of enthalpy, that is, a positive ``superadiabatic gradient" $\Delta \nabla$. It is convectively unstable according to  both the Schwarzschild  and the Ledoux criteria. With an appropriate\footnote{See \S\ref{TKE_subsection} and Eq.~\ref{ell} for an explanation of ``appropriate.''} choice of mixing length, this middle region  can be reasonably well approximated by MLT.

MLT works poorly for the bottom and top boundary layers, which have negative values of $\Delta \nabla$.
While the central region is defined by positive buoyancy, and positive enthalpy flux, outside the convective zone these quantities are zero, and in the boundaries they are negative. In MLT this is impossible because it would imply that the velocity in Eq.~\ref{mlt} is imaginary, but in Eq.~\ref{mlt_acc} merely implies buoyancy braking, hence the labels ``braking" in Fig.~\ref{qvsr}. 

\cite{zahn91} has summarized\footnote{Compare his Fig.~1 to the right braking layer in our Fig.~\ref{qvsr}; this is a nice prediction of some of the features later revealed in 3D simulations.} the issue of negative buoyancy and convective flux in connection with penetrative convection. \cite{rosner84} have discussed the overshoot at the bottom of the solar convection zone in the context of convective plumes and magnetic dynamos, and  \cite{spiegel_zahn92} have discussed this in the context of solar rotation and the tachocline. In stellar evolution theory (i.e., MLT) the existence of these braking regions is obscured by use of the Schwarzschild (or Ledoux) linear stability criterion. These braking layers are related to issues of overshoot and penetrative convection \citep{veronis63,mltz84,hurl86}. The braking layers are not a part of MLT but, as we shall see (\S\ref{braking}), arise naturally from Eq.~\ref{mlt_acc}.

\begin{figure}[h]
\figurenum{4}
\includegraphics[angle=0,scale=0.49]{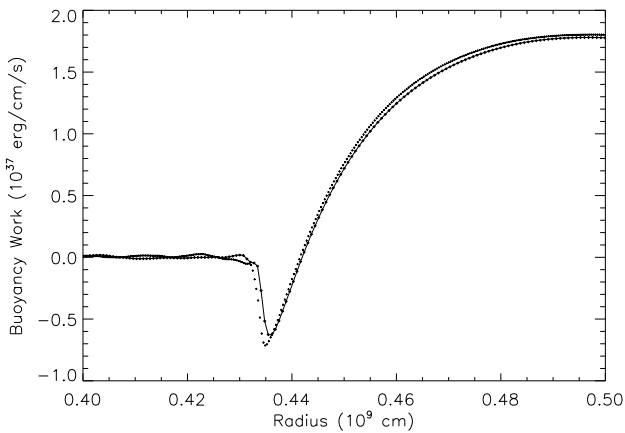}

\caption{Time-averaged buoyancy work (weighted by a factor of $4\pi r^2 \rho$)  at lower shell boundary for oxygen burning, versus radius. This shows the ``hi-res'' case of \citep{vmam13} ($768\times 512^2$) and a higher-resolution case ($1536\times1024^2$). The braking zone is indicated by negative buoyancy work at $(0.433$ to $0.445\times 10^9$ cm). Compare to Fig.~\ref{qvsr}, which shows both the upper and lower boundary for the "medium-res" case. There is a steady convergence toward a common asymptote as resolution increases, and the two cases shown here are virtually  identical, except for small variations in averaging due to differences in time step size.
}
\label{uhi-res_qvsr}
\end{figure}

 Fig.~\ref{uhi-res_qvsr} shows the inner braking zone (the region of negative buoyancy work) at $r\sim (0.433$ to $0.445\times 10^9$ cm). The ``hi-res'' case of \citealt{vmam13} ($768\times 512^2$ zones) and a still higher-resolution case of \citealt{cm15} ($1536\times1024^2$ zones) are shown. In comparison with Fig.~\ref{qvsr}, the negative ``spike" is now well-resolved. A detailed analysis of these simulations will appear elsewhere.
 The degree of numerical convergence is promising, and we conclude that such {\em braking zones are a robust feature of well-resolved simulations of neutrino-cooled stellar convection. }

\placefigure{4}
 
 The radial velocity becomes small in the braking region, while the transverse velocity extends deeper before it also becomes small. 
 The convective motion turns, and a small (mostly g-mode) wave velocity remains. The composition gradient is steeper than would be predicted by algorithmic diffusion (Eq.~\ref{eggleton}), and begins at the bottom of the braking region.
The boundary composition profiles are smooth and self-similar when time-averaged. This suggests that the turbulent spectrum has a consistent net effect on the composition profiles and on the mixing, and therefore this interface should be amenable to approximation over time-steps in 1D evolutionary calculations.

For oxygen burning, the composition gradient in the boundary layer is not well-represented by conventional turbulent diffusion theory which requires a span of many ``turbulence mean-free-paths" per density scale height \citep{harlow68} for validity\footnote{The problem is similar to that in a stellar photosphere, in which radiative diffusion must give way to radiative transfer.}. 
In MLT, the span is a fraction of a scale height (see $\Delta \ln P$ in Table~\ref{table2}) for oxygen burning. 
The small length scales are accompanied by small time scales for change, so that a steady state model may be appropriate.

\subsection{Dynamics and Braking Layers}\label{braking}

Fluid motion in a star may be separated into two fundamentally different flows \citep{ll59}: solenoidal flow (divergence free: ${\bf \nabla \cdot } \rho{\bf u }=0$) and potential flow 
(curl free: ${\bf \nabla } \times \rho {\bf u}=0$), 
which together represent the Helmholtz decomposition of an arbitrary vector field. Potential flow is associated with wave motion and solenoidal flow (vorticity) is a feature of turbulence.
A striking separation in the nature of the flow is visible at boundaries between these types of flow;  see the discussion of boundary layers in \cite{prandtl,ll59}, and 
Fig.~19 in \cite{vmam13}. This separation in types of flow is closely related to wave generation and propagation \citep{press81,pr81,gk90,gmk94}.

The structure and nature of these boundary layers 
is important for estimation of the rate at which turbulent flow moves into or from non-turbulent regions---the growth and recession of convective zones.
\cite{ma07b} had about 8 zones across the lower boundary layer for ``medium'' resolution; see also \cite{herwig14}. \cite{vmam13} had double the resolution across the convective zone (twice as many radial points), but the boundary layer became physically narrower.
Recent simulations at still higher resolution (see Fig.~\ref{uhi-res_qvsr} and \citealt{cm15}) show that the lower boundary layer has about 20 zones and the same physical depth. The computed entrainment rate may be affected by numerical viscosity, so that lower resolution simulations will give overestimates. 

The ``medium'' resolution of \cite{ma07b} was sufficient to give numerical viscosity (Reynolds number) similar to that of laboratory experiments on entrainment, but not of stars.  Coarse resolution in those simulations may have been a partial cause of the difficulties found by \cite{staritsin} in an attempt to apply the entrainment rates  of \cite{ma07b} for oxygen burning directly to main sequence stars. 
The real entrainment rates for stars should be smaller.
Another issue is that oxygen burning and hydrogen burning have very different P\'eclet numbers \citep{vmpa15}, which can affect the entrainment rate (see below).

\begin{figure}[h]
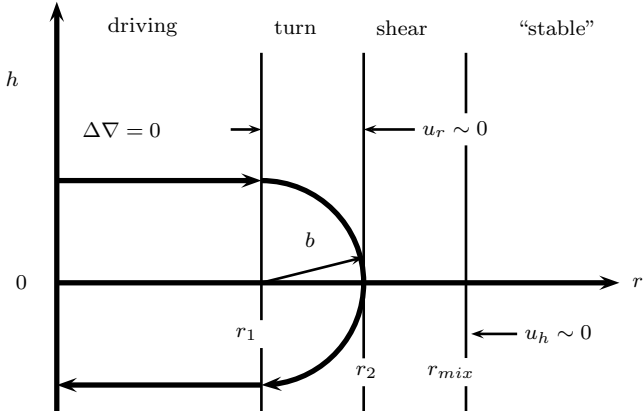

\figurenum{5}

\psset{unit=.34cm}
% picture size x is -10 to 19 y is 1 to 15
\pspicture*[](-10,1)(20,17)

% vertical (transverse) vector x is -8 and y is 0 to 14
\psline[linewidth=2pt]{->}(-8,0)(-8,17)
% horizontal axis x is -9 to 8 and y is 6
\psline[linewidth=2pt]{->}(-8,6)(14,6)
% outflow vector x is -8 to 0 y is 10
\psline[linewidth=2pt]{->}(-8,10)(0,10)
% turning arc
\psarc[linewidth=2pt]{<-}(0,6){4}{-90}{90}
% inflow vector
\psline[linewidth=2pt]{<-}(-8,2)(0,2)

%\rput*[l]{0}(-1,1){$g\beta_TH_P\Delta\nabla$}
% origin
\rput*[l]{0}(-9.6,6){0}
% radial axis label
\rput*[l]{0}(14.5,6){$r$}
% turning vector arrow
\psline[linewidth=1pt]{->}(0,6)(4,7)
% turning vector label
\rput*[l]{0}(1.7,7.7){$b$}
% label to vertical axis
\rput*[l]{0}(-10,14){$h$}
% schwarzschild label
\rput*[l]{0}(-7,12){$\Delta \nabla = 0$}
% narrow line for delta nabla label
\psline[linewidth=1pt]{->}(-1.2,12)(0,12)
% vertical line for schwarzschild
\psline[linewidth=1pt]{-}(0,15)(0,0)
% number label for schwarzschild
\rput*[l]{0}(-1.,4){$r_1$}
% vertical line for shear
\psline[linewidth=1pt]{-}(4,15)(4,0)
% vertical line for shear
\psline[linewidth=1pt]{-}(8,15)(8,0)

% number label for shear
\rput*[l]{0}(3.7,2.5){$r_2$}
% label for mix
\rput*[l]{0}(6.5,2.5){$r_{mix}$}
% label for u_r
\rput*[l]{0}(6.3,12.){$u_r \sim 0$}
% narrow line for u_r label
\psline[linewidth=1pt]{->}(6.,12)(4,12)
% label for shearing layer
%\rput*[l]{0}(6.3,9){shear}
%\rput*[l]{0}(6.3,8.5){$u_h\neq0$}
% narrow line for shear label
%\psline[linewidth=1pt]{->}(6,8.2)(4.2,6.2)
%labell for end of shearing layer
\rput*[l]{0}(10.3,4.){$u_h\sim 0$}
% narrow line for edge of shear label
\psline[linewidth=1pt]{->}(10,4.)(8.2,4.)

\rput*[l]{0}(10,16){``stable''}
\rput*[l]{0}(4.5,16){shear}
\rput*[l]{0}(0.5,16){turn}
\rput*[l]{0}(-6,16){driving}
%\rput*[l]{0}(-5,3.3){region}
\endpspicture
\caption{Simplified schematic of a convective boundary. The length $b$ corresponds to the radius of curvature needed to reverse (contain) the flow ($u_r \rightarrow - u_r$). The centrifugal acceleration is provided by pressure fluctuations (see text). The boundaries oscillate due to surface waves.
The radial direction is denoted by $r$ and the transverse by $h$. Orientation is for the top of a convection zone; the bottom may be described by appropriate reversals.}
\label{bnd2_fig}
\end{figure}

Here we construct a simple but dynamically consistent picture of a convective boundary. 
This is illustrated in Fig.~\ref{bnd2_fig}, which shows the driving, turning, shear and stable regions. 
At its most elemental level, the velocity vector must turn at boundaries; 
that is, {\em flow must turn back to stay inside the convective region}. We do {\em not} assume that ``blobs'' disappear (like MLT).
Most of the momentum is contained in the largest scales, so we focus on the average dynamics at these scales, and the simplest flow patterns.

\placefigure{5}

The magnitude of the acceleration required to turn the flow is just the centrifugal value $u^2/b$ where $b$ is the radius of the turning region and $u$ the relevant velocity. Using Eq.~\ref{mlt_acc} in the steady state limit, and taking $b \sim \Delta r \ll\ell$,
the radial component of the acceleration equation becomes
\begin{equation}
u_r \partial u_r / \partial r \sim \Delta({1 \over 2} u^2)/ \Delta r \sim {\cal B}, \label{turning}
\end{equation}
where $\cal B$  is the acceleration due to buoyancy and pressure fluctuations (Eq.~\ref{gen_acc}, and \S\ref{A_momentum}).
So far we have considered the top of a convective zone; the bottom of a convection zone behaves similarly if care is taken with signs. 

Simulations \citep{ma07b,vmam13} show 
a consistent pattern in velocity and composition structure in the boundary layers. Moving toward the boundary from the interior of the convection zone,  we find (1) the radial velocity $u_r$ decreases, (2) the pressure fluctuations $P'$ increase, and (3) the transverse velocity $u_h$ increases to a maximum and then decreases, joining on to a finite and small rms velocity due to wave motion. The transition to small rms velocity occurs at about the same point that the composition changes from being well-mixed to supporting a radial composition gradient. This pattern holds for both top and bottom boundaries.

The dynamical equations we use are derived
in Appendix~\ref{conv_append}. We use \S\ref{A_momentum}, the same quasi-steady state and thin shell ($b \ll \ell$) approximations, and choose an inertial frame in which a hydrostatic background is assumed. Near the boundary, the radial component of the acceleration is essentially just 
\begin{equation}
 {\cal B}= -\Big ({ \rho' \over \rho_0 + \rho'}\Big ) g 
-{1 \over  \rho_0 + \rho'}  \partial P'/ \partial r   . \label{radial_acc0}
\end{equation}
The buoyancy force (the first term on the RHS) is parallel to the gravity vector $\bf g$,  which is radial, and provides no transverse acceleration. 
Baryon conservation implies that this reduction in the radial velocity alone will give an increase in density (matter accumulates), which gives an increase in the pressure fluctuation $P'$ as the boundary is approached.
The two transverse components of velocity satisfy
\begin{equation}
u_h  \partial u_h / \partial h = -{1 \over \rho_0 + \rho' }  \partial P'/ \partial h . \label{transverse}
\label{trans_acc0}
\end{equation}
The transverse motion requires a transverse acceleration which is provided by a pressure excess (see also \citealt{sn89}) at the point of contact of the plume with the boundary (note the similarity to the RTI, \S\ref{rti}; and \citealt{rosner84}). 

This same pressure excess also implies a radial acceleration of the boundary, making the boundary undulate \citep{ma06,ma07b}.  In addition to the horizontal force from the pressure excess, the  buoyancy force is negative, so the net effect on the flow is to complete the turn.
The turning region has a width $b=r_2-r_1$; this material is well-mixed because it moves back into the convective region after it turns. Thus the region $r_2-r_1$ might be termed the ``over-shoot'' region, and we are discussing the dynamics of ``overshoot".

 Fig.~\ref{uhi-res_qvsr} shows our highest resolution simulation of the most demanding boundary; does this simple model of boundary dynamics work for it? The orientation is reversed for the bottom boundary, so  $r_{mix} < r_2  < r_1$ in this case.
 The steep drop in buoyancy work at $r\sim 0.433 \times 10^9\rm\ cm$ corresponds to $r_{mix}$ and the ``shear" region in Fig.~\ref{bnd2_fig},  which can maintain a composition gradient because the velocity is due to wave motion.
At the radius $r_2$, at which the radial component of the velocity is $u_r \sim 0$, the flow is transverse to the radial coordinate ($u_h \neq 0$), so there is a shear layer at this surface which will be unstable to the Kelvin-Helmholtz (KH) instability \citep{ch61}. The partial mixing layer extends to radius $r_{mix}$  (at which $u_h \sim 0$) and contains this KH layer.
The linear condition for ability of a layer to resist shear (stability against mixing) is the
``gradient'' Richardson number, $Ri >1/4$.
The Brunt-V\"ais\"al\"a frequency $N \sim 3 \rm\ s^{-1}$ is evaluated in the stable region, near the boundary, and may be sensitive to resolution. The shear velocity is $u_h \leq 0.8 \times 10^7\rm cm/s$, and from this crude estimate $r_{mix}-r_2 \sim u_h/2N \sim 10^6\rm\ cm$. This small length is consistent with the steep ``cliff'' in Fig.~\ref{uhi-res_qvsr}.

Both terms in $\cal B$ (Eq.~\ref{radial_acc0}) act to turn the flow, and are comparable in magnitude. A crude but interesting estimate follows if we take ${\cal B} \sim  g \beta_T H_P \Delta \nabla$, where the $\Delta \nabla$ is an average value over $r_2-r_1$. 
The turning radius in units of local pressure scale height is then
\begin{equation}
 b/H_P \sim \Delta r/H_P \sim \Delta({1 \over 2} u^2)\Big/  g \beta_T H_P \Delta \nabla ,\label{spike_width}
\end{equation}
which is related to the inverse of a Richardson number; compare to Eq.~\ref{Ri_gradient} and \ref{Ri_general}. Both $\Delta ({1 \over 2} u^2 )$ and $\Delta \nabla$ are negative here, giving a positive  ratio.
The use of Eq.~\ref{mlt_acc}  automatically leads to an approximate Richardson number criterion for the edge of the convective region, {\em without the need of an additional imposed boundary condition beyond the requirement that $u^2$ becomes small} (see \S\ref{imposed_bnd}).
 
 The minimum in buoyancy work at $r\sim 0.437 \times 10^9\rm\ cm$ corresponds to $r_2$, the edge of the braking region and the ``turn" in Fig.~\ref{bnd2_fig}. At $r\sim 0.443 \times 10^9\rm\ cm$ the buoyancy work becomes positive, so that this corresponds to $r_1$ and the beginning of the ``driving" region, 
 at which $\Delta \nabla$ changes sign. 
 {\em Contrary to MLT, the radius $r_1$, at which the Schwarzschild criterion is zero, is not at the boundary of zero convective motion.} 
   
How does this braking region develop a negative buoyancy? Suppose the region $r_2$ to $r_1$ is well mixed, to uniform composition and entropy. There is no braking, so convective flow is unabated to the composition gradient beginning at $r_2$. Vigorous entrainment erodes the boundary, causing a thin layer of partially mixed matter, which contains the heavier nuclei from below the oxygen burning shell. This makes the buoyancy more negative, establishing a braking layer and reducing the rate of entrainment.  The braking layer grows until the entrainment rate balances the rate of mixing into the edge of the convection zone. If the braking layer is too large, such mixing will reduce it; there is negative feedback.
The braking layer is thinner than the convective zone, so the time scale is shorter than the turnover time (\S\ref{cascade_Re}), and a quasi-steady state can be set up.
This simplistic analysis (which ignores fluctuations) indicates some of the dynamics involved with the braking layers and composition boundaries.
Further analysis with the new higher resolution simulations \citep{cm15,mmca15} is in progress.

This limiting case (``elastic collision'') is a reasonable approximation for the time averaged behavior of the oxygen burning shell \citep{ma07b}, in which radiative diffusion (and electron heat conduction) are slow;
here $\tau_{turn} \sim 0.6 \rm\  sec$, while the radiative diffusion time is 
$\tau_{diff} \sim 3 \times 10^7 \rm\ sec$. A measure of the heat lost during the turn is 
a small number ($\sim 2 \times 10^{-8}$) for oxygen burning, and is roughly the inverse of the P\'eclet number. Even within the narrow braking layer, there is little heat flow by radiative diffusion during oxygen burning.
 
 This discussion underestimates mixing because it ignores turbulent fluctuations (\S\ref{fluctuations}); larger fluctuations do more mixing than average, and mixing is irreversible. Turbulent kinetic energies fluctuate by factors $\sim 2$, so the mixing estimates should be increased accordingly. 
 Flow velocities do not go to zero at the convective boundaries, but become small and oscillatory \citep{press81,pr81,gk90,gmk94}.
As convective plumes hit the boundary, and rebound, the boundary moves in response; how elastic this is depends upon heat flow (the P\'eclet number).

This ``adiabatic''  limit breaks down as the turnover time  $\tau_{turn} \sim b/u$ approaches the radiative diffusion time for the turn 
$\tau_{diff} \sim b^2/ \lambda c$.  For larger radiation mean-free-paths, the P\'eclet number decreases.
No sharp temperature gradients can persist. This gives an ``inelastic collision''  of the flow with the boundary. 
This is the case for stars in photon-cooled stages of evolution; even with relatively large P\'eclet numbers for the whole convective region, the narrow boundary layers may still have significant energy flow by radiative diffusion. The previous discussion of the effect of excess pressure $P'$ still holds, but because of thermal diffusion $P'$ becomes increasingly dominated by density excess $\rho'$ rather than the temperature excess $T'$.

The red giant model of \cite{maxime} provides an example of a boundary layer (the bottom) in which there is significant radiative diffusion; \cite{vmam13} analyze this in detail (their \S~4.6). 
As the boundary is approached from above, the down-flows are accelerated by pressure 
dilatation.
These down-flows have an entropy deficit, so that they are heated by radiative diffusion from the surrounding material. In the braking region, compression causes  a ``hot spot" to develop. The flow is turned to a non-radial direction, and is now cooled by radiative diffusion (see Fig.~7 in \citealt{vmam13}).

Such behavior differs from that obtained by present stellar evolution algorithms. The turning of the down-flow forces the mixed region to extend  beyond that implied by the Schwarzschild criterion, and heating/cooling by radiative diffusion modifies the structure. While modest, such differences can be important for detailed models.
In compensation for such changes, a standard solar model requires less opacity to have the same convection zone depth; this implies a lower metallicity. These changes in the solar model
provide a means to reduce the disagreement with helioseismology \citep{cd11,zhang12}. \cite{dengxiong08} gave a justification for compositional smoothing, as did simulations \citep{ma07b,vmam13}. The thermal characteristics needed \citep{cd11} follow from the analysis given above, which was not designed for the solar problem, and involved no solar or stellar calibration. A more physically-correct convective boundary condition tends to improve agreement with abundances inferred from 3D stellar atmospheres \citep{asp05} and the standard solar model. 

If heat flow processes are included, the ``inelastic collision'' with the boundary allows the loss of heat so that the entropy decreases for the downward flow, enhancing the downward acceleration. This effect tends to drive motion in convective envelopes. 
Heating at the bottom also tends to drive convective flow. However, cooling at the bottom (as with URCA-shells, \citealt{wda96}) or heating at the top (downwardly entrained, burning fuel) both tend to halt the flow. Such halting processes can cause convective zones to split \citep{mocak11}.

There may be observational evidence supporting this description of boundaries of convection which are deep in stellar interiors. Detection of g-mode pulsations in subdwarf B (sdB) stars allows an asteroseismic estimation of the size of the He-burning cores, which are significantly larger than predicted by the Schwarzschild criterion and standard stellar evolution theory (see \citealt{sga15} for discussion and references). Similar issues apparently are general for core helium burning stars observed by Kepler \citep{keplerRG,ccl14}.

Finally, the origin ($r=0$), in a 1D stellar evolutionary code using MLT, is a boundary as well. The use of Eq.~\ref{mlt}  with adequate zoning implies that the convective velocity becomes very small due to symmetry (derivatives go to zero at the origin). {\em This is a false braking layer caused by MLT being a local theory.} Use of Eq.~\ref{gen_acc} allows flow through the origin  provided that a counter flow gives conservation of linear momentum (e.g., a 
{\bf toroidal
}
 roll). At the origin in a turbulent convective core,  this projects onto 1D as  a finite rms velocity, with a zero radial gradient. MLT has problems with velocity at $r=0$.

\section{Summary}\label{summary}

We have brought more precision to the discussion of stellar convection by the use of 3D simulations of sufficient resolution to exhibit truly turbulent flow and boundary layers.  The price paid is that we must replace the unresolved turbulent cascade by Kolmogorov theory (ILES approximation), and the chaotic behavior of an integral scale roll of Lorenz by a steady-state average. {\em We use RANS averaging to make 3D simulation data concise, and use 3D simulations to give RANS closure.} Solution of the RANS equations, using only the significant terms \citep{mocak14}, is the full 321D procedure.

This approach gives us a quantitative and precise foundation, based upon turbulent solutions of the equations of fluid dynamics. These numerical solutions have numerical limitations, which we have discussed.
We find that the actual sub-grid dissipation in our simulations is automatically well approximated by the Kolmogorov four-fifths law.

As a simpler first step, which addresses some of the worst errors of MLT, we focus on the acceleration equation for the turbulent velocity. This makes the theory non-local, time dependent, and produces boundary layers. It is almost identical to the equation developed from experimental study of the Rayleigh-Taylor instability (RTI), indicating a close connection with plume models of convection; simulations also suggest this connection directly.  Further development would entail use of RANS analysis to better deal with turbulent fluctuations (\S\ref{fluctuations} and \ref{TKE_subsection}).

Even within the framework of the simple acceleration equation, there are several indications of how current practices in stellar evolution could be improved. The least drastic change involves diffusion: artificial diffusion (\S\ref{eggleton_diff}) should be used with caution in situations in which real diffusion (\S\ref{michaud}) operates, because of distortion of the gradients which drive real diffusion (both artificial and real diffusion have second-order spatial derivatives). The discussion in \S\ref{braking} gives a more realistic way to treat ``overshooting'', and at the same time, removes the need for an imposed boundary condition (Schwarzschild, Ledoux, or Richardson; 
\S\ref{imposed_bnd}). The fluctuations in pressure discussed in \S\ref{braking} will cause wave motion which will drive mixing in  semi-convective regions on a dynamical timescale, far faster than the thermal timescale conventionally used (e.g., \cite{leef85}; see \S\ref{semiconv}).

 For use in stellar evolution this approach requires one more differential equation (for velocity, in addition to the traditional four, e.g., $r$, $L$, $T$, and $\rho$) and additional coupling terms in the usual stellar evolution differential equations (turbulent heating in the energy equation, and ram pressure in the hydrostatic equation). The additional demand upon computational resources is not large.
 We use the convective flow velocity $\bf u$ and the super-adiabatic excess $\Delta \nabla$ as separate variables, reflecting the fact that they have different correlation lengths \citep{ma07b}.
We check that the simplified dynamic model does capture the numerical results of 3D as expressed in the RANS formulation. {\em This approach is not calibrated to astronomical data, but predictive, being based on simulations and laboratory experiment.
The simple 321D approach includes the Kolmogorov-Richardson turbulent cascade,
and allows connections to past and future numerical simulations as a natural consequence. }

\subsection{The future}

The enormous simplification, from 3D turbulent simulations requiring terabytes of storage down to a single additional ordinary differential  equation (e.g., Eq.~\ref{gen_acc}), means that much is missing. For some applications the missing items may be important. One might use the RANS equations directly in a stellar evolutionary code, with 3D simulations to guide closure \citep{mocak14}. We have presented a step toward that goal. Alternatively, one might add to the simple 321D as needed, using new models guided by RANS results. Probably both paths should be followed, given the complexity of the problem.

\subsubsection{321D algorithms}
We have refrained from offering detailed algorithms because we believe that there may be a variety of useful ones,  tailored for existing stellar evolution codes, and to be modified by developing insight. This is not a finished subject. A skeleton algorithm should include:
\begin{enumerate}
\item  velocity from an acceleration equation (Eq.~\ref{gen_acc}, \S\ref{dynamics}),
\item boundary physics: turning, damping, mixing and shear (\S\ref{braking}),
\item fluxes of enthalpy and composition (\S\ref{nonuniformY} and \S\ref{braking}),
\item non-locality in velocity: turbulent kinetic energy flux and ram pressure (\S\ref{dynamics}), and
\item turbulent heating of background by Kolmogorov cascade (Eq.~\ref{kolmog_eps}).
\end{enumerate}
Our first priority is to implement these ideas in stellar evolution codes. We are currently  testing in TYCHO (\citealt{liebert13}), and
plan to migrate to MESA \citep{mesa,mesa13}, MONSTAR (\citealt{monstar08,monstar10}), GENEC (\citealt{sjones15}), and
FRANEC (\citealt{franec13}).
We will gladly help with implementations in other codes.

\subsubsection{Further simulations}
New simulations to better quantify the boundary physics are in progress (\citealt{cm15}; \citealt{andrea}).
This approach, unlike MLT, is generalizable in principle to include rotation and MHD  \citep{maeder,mm00}
because it starts with full 3D equations. For example, rotational terms are implicit in the vector form of Eq.~\ref{gen_acc}; see also \cite{balbus09,feather15}. 

\subsection{Implications}
Because of the fundamental importance of convection in stellar evolution theory, a replacement for MLT will have implications for many areas throughout astronomy and astrophysics. A few of the most striking are:

\subsubsection{Helioseismology}
Convective boundaries with low P\'eclet number will be smoother, which reduces the disagreement between helioseismology and solar model predictions; see \citealt{cd11,zhang12} and \S\ref{sect3}. 

The corrected boundary conditions for convection will place the composition gradient further beyond the Schwarzschild zero condition (\S\ref{braking}), requiring a lower opacity below the mixing boundary to get an acceptable solar model. This may be attained by a lower metallicity, which will reduce the disagreement between solar models, and solar abundances determined from 3D atmospheres \citep{asp05}. {\em The combination of  these two corrections will shift the standard solar model problem toward the Asplund abundances. }

\subsubsection{Asteroseismology}
These modifications beyond MLT bear on many discrepancies between asteroseismology and stellar evolution theory. 
Some examples: application of better convective boundary physics will produce larger He burning cores in sdB stars, and reduce the large discrepancy between the asteroseismology determination of core sizes and stellar models \citep{Charpinet1997,vangrootel10,Bloemen2014,sga15}. 
Similar issues apparently are general for core helium burning stars observed by Kepler \citep{keplerRG}.
The discrepancy in mixed modes in normal CHeB (``red clump'') stars 
\citep{bildstens12,montalban13,stello13,ccl14} will be affected.

\subsubsection{Convective boundaries, nucleosynthesis yields and pre-supernovae}
The nature of convective boundaries is affected by radiative diffusion, so that they differ for neutrino-cooled stages of nuclear burning. Calibration of convection for late stages, from stages dominated by photon-cooling, requires re-evaluation. Detailed estimates of stellar nucleosynthesis and stellar structure based upon an algorithmic diffusion scenario (e.g., \citealt{ww95,whw02}) are not confirmed, and require re-examination. 

While the general features of nucleosynthesis yields are robust \citep{wda96}, detailed abundances depend upon details of mixing and convection.
Nucleosynthesis from lower mass stars is also affected: asymptotic giant branch (AGB) stars do not have a third dredge up without ``overshoot'', which is a convective boundary problem. This dredge up is crucial for s-process nucleosynthesis (it provides a neutron source, \citealt{lattanzio14}).

Driven by neutrino cooling, nuclear burning  in stars prior to core collapse is vigorous, and in turn drives vigorous convection.
Convective velocities increase as evolution proceeds.
The nuclear energy generation is, on average, in balance with the turbulent dissipation at the Kolmogorov scale, so
$   \epsilon_{nuc} \sim u^3 / \ell $, which relates the nuclear energy generation rate, the average convective velocity, and the depth of the convective zone. Velocity fluctuations are large \citep{ma07b}. Supernova progenitor models which are 1D can represent average properties, such as convective speed,  but not the amplitude and phase of the (large) fluctuations of those properties. 
Realistic progenitor models should be dynamic and 3D \citep{am11a,am11b} if they are to be used for accurate core collapse simulations.

\subsubsection{Core collapse}
 The size and structure of progenitor cores affects the possibility of producing explosions in core collapse simulations \citep{cott13,wda14}. The predicted size and structure of such cores depends upon the physics of convection used in the stellar evolution codes. Detailed scenarios for pre-supernova  structure, collapse and explosion, such as found in \cite{whw02} for example, are not robust, and may require revision when better treatments of mixing are applied.
 The validity of calibrating neutrino cooled convection on photon cooled stages of evolution is questionable due to the large difference in P\'eclet number.
 Even the size of the He core is uncertain with present algorithms \citep{langer91,langer12}, and will be affected by better treatment of convection and convective boundaries.
 The theoretical approach to turbulence used above can also be applied to the core collapse process itself
\citep{mm11}, giving insight even for 3D simulations which are presently under-resolved due to computational limitations.

\begin{acknowledgements}
This work was supported in part by NSF 0708871, 1107445,
NASA NNX08AH19G at the University of Arizona, and
 by Australian Research Council grants DP1095368 and DP120101815 (J. Lattanzio, P. I.) at Monash University, Clayton, Australia,
and by the European Research Council through grant ERC-AdG No. 341157-COCO2CASA.
This work used the Extreme Science and Engineering Discovery Environment (XSEDE), which is supported by National Science Foundation grant number OCI-1053575, and made use of ORNL/Kraken and  TACC/Stampede.
This work was supported in part by resources provided by the Pawsey Supercomputing Centre with funding from the Australian Government and the Government of Western Australia, and
 through the National Computational Infrastructure under the National Computational Merit Allocation Scheme.
This work was supported in part by the National Science Foundation under Grant No. PHYS-1066293 and the hospitality of the Aspen Center for Physics.
We wish to thank Alvio Renzini for asking WDA (repeatedly) ``why does MLT work?",  Vitorio Canuto for helpful hints, and 
Marco Limongi, Alessando Chieffi, Norman Murray, Bill Paxton and Stan Owocki for helpful and encouraging discussions.
One of us (WDA) wishes to thank Prof. Remo Ruffini of ICRAnet, 
and Prof. Lars Bildsten of the Kavli Institute of Theoretical Physics, for their hospitality and support.
We wish to thank an anonymous referee for extensive comments which helped improve the paper.
\end{acknowledgements}

\appendix

\section{The Convection Equations}\label{conv_append}

We develop the fluid equations in an inertial frame \citep{ll59}. We begin with a general formulation, and transition to a specifically spherical ($r,\theta,\phi$) choice of coordinates for application to stars.
We will decompose variables into a background part and a fluctuating part, e.g., 
for pressure $P=P_0+P'$. Our procedure is chosen for stars in which the background is hydrostatic and spherically symmetric, so that $ \nabla P_0 = - \rho_0 {\bf g} = - {\bf g}/V_0$. 

\subsection{Baryon Conservation}

The vector form of the continuity equation \citep{ll59} is
\begin{equation}
\partial \rho/\partial t + {\bf \nabla \cdot}\rho{\bf u} = 0, \label{continuity}
\end{equation}
where $\rho$ is the mass density and $\bf u$ is the fluid velocity.
In the incompressible limit, for a steady flow, the net flux of mass into a region equals the mass flux out. In thin boundary layer, perpendicular to the radial direction $r$, the average velocities must satisfy
 \begin{equation}
\partial u_r /\partial r = - 2 \partial u_h / \partial h,
\end{equation}
where $h$ is either of the symmetric transverse coordinates (i.e., locally cartesian), to avoid changing the density (as seen in the Eulerian frame).

\cite{vmam13} show (their Eq.~28), that for fluctuations against a steady background,
\begin{equation}
{\bf \nabla \cdot u'} = {u_r' \over H_\rho},
\end{equation}
where $H_\rho$ is the density scale height, and $u_r'$ is the radial component of the velocity fluctuation.
This approaches zero (the incompressible limit) for shallow, subsonic convection (large density scale height and small radial velocity mach number, $u_r' \ll s$,  where $s$ is the sound speed). This velocity ``dilatation'' is due to the vertical motion in the background stratification and becomes an important component in convective driving in deep convection zones \citep{vmam13}. Notice that rising plumes ($u_r >0$) expand and falling plumes contract \citep{sn89,ma10}.

\subsection{Momentum Conservation}\label{A_momentum}

The vector acceleration equation (Eq.~\ref{gen_acc}) is
\begin{equation}
\partial {\bf u}/ \partial t + {\bf (u \cdot \nabla) u} = {\cal B} -{\bf u}/\tau \label{NSK}
\end{equation}
where  $\bf u$ is the velocity, $\tau = |u|/\ell_d$ with $\ell_d$ is the Kolmogorov damping length, and the variable $\cal B$ is defined as in \S\ref{dynamics}. If
\begin{equation}
{\cal B} = -{1 \over \rho} {\bf \nabla} P - {\bf g} ,
\end{equation}
where $P$ is pressure and $\bf g$ is gravitational acceleration,
then Eq.~\ref{NSK} is a Navier-Stokes description of the largest scales of turbulence, with a simplified damping term which is consistent with \cite{kolmg}.
Note that the usual formulation of hydrostatic equilibrium in stellar evolution theory is some variant of the condition ${\cal B }=0$.
Projecting Eq.~\ref{NSK} onto the radial coordinate, we have
\begin{equation}
 \partial u_r /\partial t + u_r\partial u_r/\partial r= -  {1 \over \rho} \partial P /\partial r - g  -u_r/\tau .
\end{equation}

The full equations in spherical coordinates are shown in \S{15}, \cite{ll59} (see also \citealt{mm84} for a detailed discussion), with the bare viscosity terms rather than Komogorov's expression for integration of the turbulent cascade. 
In tensor form the momentum equation is
\begin{eqnarray}
\partial u_i / \partial t + u_k \partial u_i /\partial x_k = -{ 1 \over \rho}\partial P /\partial x_i -g_i \nonumber\\
+{ 1 \over \rho}{\partial \over \partial x_k} \Big [ \eta \Big ( {\partial u_i \over \partial x_k} + {\partial u_k \over \partial x_i} -{2 \over 3}\delta_{ik}{\partial u_l \over \partial x_l} \Big ) \Big] + {\partial \over \partial x_i} \Big (\zeta {\partial u_l \over \partial x_l }\Big ). \label{tensor-NS}
\end{eqnarray}
Kolmogorov's four-fifths law \citep{frisch} states an amazing simplification, that integration over the turbulent cascade reduces the last term  in Eq.~\ref{tensor-NS} to $-{\bf u}/\tau$ (Eq.~\ref{NSK}) on average, ignoring boundary effects (see \S\ref{sect3}).

To illustrate how turning happens at boundaries, it is sufficient to consider
the simpler case of flows with $\theta$ and $\phi$ length scales small compared to $r$, so the transverse dimensions  are quasi-cartesian (the inertial terms in $1/r$ are neglected; for convective cores, the more cumbersome full equations are needed because $r$ cannot be  large near the origin).
Then the two transverse components are symmetric in this approximation and satisfy
\begin{equation}
\partial u_h/ \partial t +  u_h  \partial u_h / \partial h = -{1 \over \rho}  \partial P/ \partial h  -u_h/\tau,
\end{equation}
where $dh$ is $rd\theta$ or $r \sin \theta d\phi$.
We consider finite fluctuations about a static background, so that we substitute $\rho = \rho_0 + \rho'$ and $P = P_0 +P'$. We ignore variations in $g$ (the Cowling approximation, \citealt{cox80}).
Using $-\partial P_0 /\partial r = \rho_0 g$, the radial equation becomes
\begin{equation}
\partial u_r/ \partial t +  u_r  \partial u_r / \partial r = -\Big ({ \rho' \over \rho_0 + \rho'}\Big ) g -{1 \over  \rho_0 +\rho'}  \partial P'/ \partial r  -u_r/\tau .\label{Bradial_eq}
\end{equation}
Convection is often described using only the buoyancy term; the pressure fluctuations are taken to be small, of order the mach number squared. However, near boundaries the pressure fluctuations provide the tangential acceleration which is necessary to turn the flow, and should not be neglected (see \citealt{aake85}).
The buoyancy term acts through the density fluctuation $\rho'$, and only in the direction parallel to the gravity vector. The transverse equation is
\begin{equation}
\partial u_h/ \partial t +  u_h  \partial u_h / \partial h = -{1 \over \rho_0 + \rho'}  \partial P'/ \partial h  -u_h/\tau . \label{Btransverse_eq}
\end{equation}
Note that the radial and transverse equations are coupled primarily by the pressure fluctuation term $P'$, but also by $u/\tau$, because
$ \tau = \ell_d / |u|$ where $|u|^2 = u^2 = u_r^2 + 2u_h^2$ (turbulence damps regardless of orientation of the large scale flow).
The fluctuating pressure near convective boundaries insures the generation of waves.

\subsection{Energy Conservation}
Following \cite{ll59}, \S6, the equation of energy conservation is
\begin{equation}
{\partial \over \partial t} \Big ( {1\over 2} \rho u^2 + \rho E +\rho \phi \Big ) = - {\bf \nabla \cdot} \Big [ \rho{\bf u} ( {1\over 2} u^2 + W +\phi ) \Big ] + T{ \partial \rho S \over \partial t }, \label{energy_eq}
\end{equation}
where $\phi$ is the gravitational potential and ${\bf g}= - \nabla \phi$. If taken to both the steady  state and adiabatic limits, this becomes  the Bernoulli equation \citep{ll59}.
The entropy change equation may be written as
\begin{equation}
T{ \partial \rho S \over \partial t } = \rho \epsilon_{nuc} + \rho \epsilon_{visc} - {\bf \ F_{rad} }, \label{entropy_term}
\end{equation}
where $\epsilon_{nuc}$ is the net heating from nuclear and neutrino reactions, $\rho \epsilon_{visc} $ is the Navier-Stokes viscous heating term as modified by Kolmogorov's four-fifth's law (see Eq.~\ref{kolmog_eps}, \ref{NSK} and \ref{tensor-NS}), and $F_{rad}$ is the energy flux due to  radiative diffusion. The viscous term is missing from MLT and the Euler equation.
Most of the turbulent kinetic energy resides in the largest (integral) scale,
while turbulent heating occurs at the small (Kolmogorov) scale. Then $\epsilon_{turb} = {\bf u \cdot u} |u|/\ell_d$ is the Kolmogorov heating from the turbulent cascade, and 
$T{ \partial \rho S / \partial t }$, 
$\rho \epsilon_{nuc}$ and 
$F_{rad}$ are now the appropriate RANS averages \citep{vmam13}. 
One requirement for Bernoulli's equation to be valid, as assumed in \cite{pasetto14} (see \S\ref{pasetto}), is that the RHS of Eq.~\ref{entropy_term} must be zero (\citealt{ll59}, Ch.~I). This is found not to be generally true, either in the 3D simulations \citep{vmam13,mocak14}, or experimentally in turbulent flows \citep{tennekes,davidson}. Heating is an essential feature of 3D turbulence, which converts large scale, ordered velocities to disordered ones.

\

\clearpage

\end{document}